\documentclass[reprint,superscriptaddress,amsmath,amssymb,aps,prl]{revtex4-2}

\usepackage{amsmath, amssymb, mathtools, physics}
\usepackage{graphicx}
\usepackage{calc} 
\usepackage{dcolumn}
\usepackage{bm}
\usepackage[colorlinks]{hyperref}
\usepackage{tikz}
\usepackage{booktabs}
\usepackage[caption=false]{subfig}
\usetikzlibrary{arrows.meta,decorations.pathmorphing,positioning}

\begin{document}

\preprint{APS/123-QED}

\title{Noise-Induced Limits on Responsivity and SNR for Nonlinear Exceptional Point Sensing}

\author{Todd~Darcie}
 \affiliation{Department of Electrical and Computer Engineering, \\University of Toronto}%
 \email{todd.darcie@mail.utoronto.ca}
\author{J. Stewart Aitchison}%

\affiliation{%
  Department of Electrical and Computer Engineering, \\University of Toronto}
  
\date{May 25, 2025} 
\begin{abstract}
  
Exceptional points (EPs) have been suggested for ultra‑sensitive sensing because the eigenfrequency splitting grows as the nth-root of a perturbation, suggesting divergent responsivity. In ideal linear devices, however, this responsivity gain is reconciled by a matching divergence in the quantum shot-noise floor, so the net signal-to-noise ratio remains unchanged. Recent work has extended this argument to nonlinear devices, such as above‑threshold lasers, predicting other divergences at an EP that is shifted by the interplay of noise and saturation effects. Here we analyze a system of two coupled saturable resonators and show analytically that a self‑consistent treatment of fluctuation dynamics removes these divergences entirely. Islands of instability arise in the parameter space surrounding the EP due to the coupling of phase noise into the amplitude dynamics, dictating a maximum responsivity and maximum noise that can be experimentally observed. Stochastic Langevin simulations of the full nonlinear system corroborate our analytical results down to zero detuning. 

\end{abstract}


\maketitle
Exceptional points (EPs) in non‑Hermitian systems are spectral singularities at which two or more eigenvectors and their eigenvalues coalesce, reducing the system’s effective dimensionality and endowing it with an unusual topology \cite{Feng2017,ElGanainy2018,Miri2019}.  A hallmark of this topology is the nonlinear eigenfrequency response: for a second‑order EP the splitting of the originally degenerate modes scales as $\Delta\lambda\propto\sqrt{\varepsilon}$, where $\varepsilon$ is a small external perturbation\,\cite{Kato2012}. The effective order of the EP can also be increased through the introduction of nonlinearity, leading to further responsivity enhancements \cite{Darcie2025, Bai2023a}. The resulting formal divergence of the slope $\partial\Delta\lambda/\partial\varepsilon$ has fueled proposals for ``hypersensitive" EP‑based sensors across optics, microwaves, and acoustics\,\cite{Chen2017,Hodaei2017,Wiersig2016,Wiersig2020}.

Linear coupled‑mode analyses, however, reveal a simultaneous divergence of the shot‑noise level that precisely cancels the gain in responsivity, leaving the signal‑to‑noise ratio (SNR) unchanged at the EP \cite{Lau2018,langbein2018, Chen2019, Loughlin2024}. This has also been confirmed experimentally \cite{Wang2020}. The noise enhancement, quantified by the Petermann factor of the linearized Hamiltonian \cite{Petermann1979}, arises from the same non-orthogonality of eigenmodes that leads to the enhanced response \cite{Wiersig2016}. Extending the discussion to nonlinear devices, previous theoretical work has either argued that saturation merely displaces the EP in parameter space while preserving the twin divergences of responsivity and noise \cite{Zheng2025}, or ignored the impact of higher-order fluctuations entirely \cite{Smith2022}.  In these treatments the fluctuation dynamics enter only as an additive noise source, and their feedback on the mean fields is neglected. 

In this work we revisit the problem using two conservatively coupled saturable resonators as a representative nonlinear platform.  By treating amplitude–phase fluctuations self‑consistently within a Langevin framework, we show analytically that both the responsivity and noise floor remain finite at the shifted EP; below a critical perturbation strength the response crosses over to the ordinary linear regime. This happens because the Hamiltonian governing the fluctuations, rather than that of the mean fields, must be tuned to an EP to achieve a true $n^{\text{th}}$--root response in the observable output frequency. Stochastic simulations of the complete nonlinear equations corroborate our analytic results and illustrate the shortcomings of prior approaches in the low detuning regime, including at the location of the EP. Furthermore, we find that operating precisely at the EP of this fluctuation Hamiltonian is not feasible due to the coupling of enhanced phase noise into the amplitude subsystem, which leads to instability. This effect imposes an upper limit on the responsivity of any EP sensor subject to additive noise and nonlinearity.   

\begin{figure}[h]
    \centering
    \includegraphics[]{  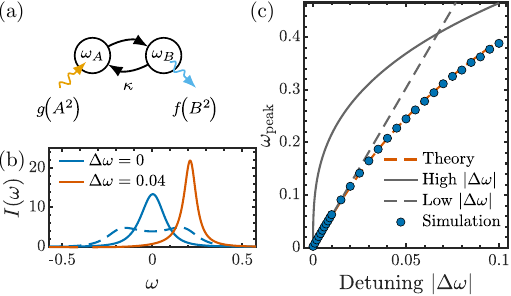}
    \caption{(a) Schematic of an resonator with nonlinear gain, which depends on amplitude $A$, coupled to a resonator with nonlinear loss, which depends on $B$.  (b) Averaged power spectra $I(\omega) = \langle |\Psi(\omega)|^2\rangle$, calculated from $\Psi (t) = A(t) e^{i\theta_1(t)}$, over 64 independent simulations at the location of the noiseless EP ($\Delta\omega=0$, $\gamma_2=\kappa=1$, dashed blue), and at the shifted EP ($\Delta\omega=0$, $\gamma_2=1.088$, solid blue), and detuned from the shifted EP (solid orange).  (c) Peak frequency near the shifted EP. The blue dots are extracted from the averaged spectra via curve-fitting; the orange dashes show the theoretical predictions from Eqs.~\eqref{theta1_dot} and \eqref{expansion_EP}. The dashed gray line shows the small detuning approximation Eq.~\eqref{thetadotweak1}, where the response is linear. The solid gray lines show the frequency response in the opposite limit, where the noise is insignificant. The other parameters for (b) and (c) are $g_0 = 4, f_0 = 0, \gamma_1 = 1,$ and $D_1 = 0.01$.}
    \label{fig:spectrum}
\end{figure}

Our generalized example system consists of two conservatively coupled resonators, as shown in Fig.~\ref{fig:spectrum}(a). One resonator is subject to a nonlinear gain $g$, and the other a nonlinear loss $f$. The resonators have natural frequencies $\omega_A = \omega_0 + \Delta \omega$ and $\omega_B = \omega_0$, where $\Delta \omega$ is a detuning that is introduced between them. It evolves according to the polar Langevin equations \cite{Zhang2018}

\begin{align}
\dot{A} &= g (A^2) A - \kappa B\sin(\theta_2-\theta_1) + \xi_A, \label{Adot} \\
\dot{B} &= -f(B^2) B + \kappa A\sin(\theta_2-\theta_1) + \xi_B, \label{Bdot} \\
\dot{\theta_1} &= \omega_A - \frac{\kappa B}{A}\cos(\theta_2-\theta_1) + \xi_{\theta_1}, \label{theta1_dot}\\
\dot{\theta_2} &= \omega_B - \frac{\kappa A}{B}\cos(\theta_2-\theta_1) + \xi_{\theta_2}\label{theta2_dot}, 
\end{align}

where $A, B$ are the respective (classical) $c$-number mode amplitudes for the two oscillators, $\theta_1, \theta_2$, are the phases, $\kappa$ is the coupling strength, and $\xi_j$ are zero mean white noise terms which satisfy $\langle\xi_{\mu}(t)\xi_{\nu}(t')\rangle=D_{\mu\nu}\delta(t-t')$. The non-vanishing diffusion matrix elements in this basis are $ D_{AA} = \frac{D_{1}}{2}$, $ D_{BB} = \frac{D_{2}}{2}$, $D_{\theta_1 \theta_1} = D_1/ 2\langle A^2 \rangle$, and $D_{\theta_2 \theta_2} = D_2/ 2\langle B^2 \rangle$. Eliminating the common phase rotation at $\omega_0$, Eqs.~\eqref{theta1_dot}-\eqref{theta2_dot} combine to yield 

\begin{align}
\dot{\varphi} &= \Delta \omega - \kappa\left(\frac{A}{B} - \frac{B}{A}\right)\cos\varphi + \xi_{\varphi}, \label{varphi_dot}
\end{align}

where $\varphi = \theta_1 - \theta_2$ is the relative phase between the two resonators, and $\xi_\varphi = \xi_{\theta_1} - \xi_{\theta_2}$ is described by the diffusion constant  $D_{\varphi \varphi} = D_{\theta_1 \theta_1} + D_{\theta_2 \theta_2} $. The detuning is defined as $\Delta\omega = \omega_A - \omega_B$. In this work, we assume that the gain and loss have both saturable and linear (passive) components such that $ g = g_0/(1 + A^2) - \gamma_1$ and $f = f_0/(1 + B^2) + \gamma_2$. However, our results generalize to any analytic functional form for $g, f$. The thresholds, stability, and frequency response of this system are thoroughly discussed in Ref.~\cite{Darcie2025}. In a hypothetical system with the noise terms $\xi_i$ turned off, we found that the system exhibits an EP with $ g = f = \kappa$. The frequency response near that operating point is proportional to $\Delta\omega^{1/3}$, implying a diverging responsivity 

\begin{equation}
    \mathcal{S} \equiv |\partial \omega /\partial \Delta\omega|^2 \propto \Delta\omega^{-4/3}.
\end{equation}

Despite using only two coupled resonators, this scaling is characteristic of a linear EP3 \cite{Hodaei2017}. Several studies of similar nonlinear systems have sought to quantify the noise level by looking at the eigen-states of the instantaneous  Hamiltonian 

\begin{equation}
H(t)  =
\begin{pmatrix}
\omega_0 + \Delta \omega + i\,g & \kappa \\[1mm]
\kappa & \omega_0 -\,i\,f
\end{pmatrix}, \label{H}
\end{equation}

which are used to calculate the Petermann excess noise factor \cite{Bai2023a, H.Haus1985, Siegman1989}. Recent experiments have shown that this treatment approximately accounts for the scaling of noise near linear EPs \cite{Wang2020}. However, the Petermann factor was initially derived for linear systems \cite{Petermann1979}, and fails to generalize to nonlinear systems \cite{Lee2000, Smith2022}. To demonstrate this for our system, we expand $A(t)=A_0+\delta A(t)$, and likewise for $B(t)$ and $\varphi(t)$. Here the fluctuation terms have zero mean, such that $\langle A \rangle = A_0$. The evolution of the fluctuation terms are described by 
\begin{equation}
    \mathbf u
=\bigl(\,\delta A,\;\delta B,\;\delta\varphi\,\bigr)^{\mathsf T},
\qquad
\dot{\mathbf u}=J\,\mathbf u + \Xi, \label{fluctuation_dynamics}
\end{equation}
where  $\Xi = [\xi_A, \xi_B, \xi_\varphi]^{T}$ is the vector of noise variables, and the deterministic Jacobian is 
\begin{equation}
    J =
\begin{pmatrix}
\Gamma_{A} 
  & -\,\kappa\,\sin\varphi_{0} 
  & \kappa\,B_{0}\,\cos\varphi_{0}\\[6pt]
\kappa\,\sin\varphi_{0} 
  & -\,\Gamma_{B} 
  & -\,\kappa\,A_{0}\,\cos\varphi_{0}\\[6pt]
C_{A} 
  & C_{B} 
  & -\,\lambda_{\varphi}
\end{pmatrix},
\label{Jacobian_full}
\end{equation}
where he have defined linearized gain and loss
\begin{align}
\Gamma_{A}&= g(A_{0}^{2})+2A_{0}^{2}g'(A_{0}^{2}),\\
\Gamma_{B} &= f(B_{0}^{2})+2B_{0}^{2}f'(B_{0}^{2}),
\end{align}
as well as $C_{A}= \kappa A_0^{-1}\cos\varphi_{0}(r_{0}+r_{0}^{-1})$,  $
C_{B}= \kappa B_0^{-1} \cos\varphi_{0}(r_{0}+r_{0}^{-1})$, $\lambda_{\varphi} = \kappa (r_0 - r_0^{-1}) \sin\varphi_0$ , and $r_0 = A_0/B_0$. In Sec.~\ref{sec:noise_scaling_weak} of the Supplemental Material, we show that this treatment predicts a faster noise scaling than that of the Petermann result; the true noise floor has the same $\Delta\omega^{-4/3}$ divergence as the responsivity, resulting in no increased signal-to-noise ratio (SNR) \cite{supplement}. 

While this result follows from keeping only first order terms in the fluctuations, recent work has shown that keeping second-order fluctuations results in a nonzero bias that shifts the mean values of the nonlinear gain and loss, and therefore moves the system away from the EP \cite{Zheng2025}. This bias arises from the fact that the effective (average) nonlinear gain  $\langle g(A) \rangle \neq g(\langle A \rangle) = g(A_0)$ once second order moments are included. Using a standard Taylor series expansion, the effective gain and loss instead take the forms

\begin{align}
g_{\mathrm{eff}} 
&= g(A_0^{2})
+ \bigl[g'(A_0^{2}) + 2A_0^{2}g''(A_0^{2})\bigr]\;
      \langle\delta A^{2}\rangle \label{g_eff},\\
      f_{\mathrm{eff}}
&= f(B_0^{2})
+ \bigl[f'(B_0^{2}) + 2B_0^{2}f''(B_0^{2})\bigr]\;
      \langle\delta B^{2}\rangle \label{f_eff}.
\end{align}

To find location of the ``shifted" EP, the authors define an effective Hamiltonian $H_{\mathrm{eff}} \equiv \langle H(t) \rangle$ for the macroscopic fields by substituting $g$ and $f$ with $g_{\mathrm{eff}}$ and $f_{\mathrm{eff}}$ in Eq.~\eqref{H}. The eigenvalues and eigenvectors of $\langle H \rangle$ coalesce when $\Delta\omega = 0$ and $g_{\mathrm{eff}} + f_{\mathrm{eff}} = 2 \kappa$, which are taken as the generalization of the deterministic EP conditions. We outline our procedure for reaching these shifted EP conditions in Sec.~\ref{sec:noise_adjusted_EPs} of the Supplemental Material \cite{supplement}. The power spectra before and after this tuning step are also shown in Fig.~\ref{fig:spectrum}(b). In the former, a frequency splitting arises at zero detuning due to the interplay of noise and nonlinearity. From the eigenvectors of $H_{\mathrm{eff}}$, it is straightforward to show that the phase difference between the two resonators is equal to $\pi/2$ when the shifted EP conditions are satisfied, consistent with the deterministic treatment \cite{Darcie2025}. However, since $g_{\mathrm{eff}}$ and $f_{\mathrm{eff}}$ receive different noise corrections, we cannot generally say that $A_0 = B_0$ at this point, as we would in the deterministic system.  

We find that this amplitude asymmetry has significant implications for the dynamics of both the fluctuations and the macroscopic fields in the vicinity of this operating point. To illustrate this, we begin by initially neglecting amplitude fluctuations, expanding $\dot{\varphi}$ (Eq.~\eqref{varphi_dot}) to second order, and taking a time-average, which gives

\begin{equation}
\cos\varphi_0
=\frac{\Delta\omega}{\kappa \Delta r \left(1-\dfrac{1}{2}\sigma_{\varphi}^2\right)}, \label{cos_varphi} 
\end{equation}

where we have defined the amplitude asymmetry factor $\Delta r = r_0 - r_0^{-1}$. When $J$ is stable, the variance $\sigma_{\varphi}^2 \equiv \langle \delta\varphi^2 \rangle$ can be uniquely determined from Eq.~\eqref{Jacobian_full} as 

\begin{equation}
\sigma_{\varphi}^2
= \frac{D_{\varphi\varphi}}
          {2\,\kappa \Delta r\sin\varphi_0}\label{sigma_varphi}.
\end{equation}

Together, Eqs.~\eqref{cos_varphi}-\eqref{sigma_varphi} provide two relations between the amplitude asymmetry and the relative phase $\varphi_0$. Applying the identity $\sin^2\varphi_0+\cos^2\varphi_0=1$ and keeping terms up to $\sigma_{\varphi}^4$ gives the detuning-dependent asymmetry factor 

\begin{equation}
\Delta r
= 
  \,\sqrt{\,\frac{\Delta\omega^2}{\kappa^2}
       + \Delta r_0^2
       } \label{Delta r},
\end{equation}

where $\Delta r_0 = D_{\varphi\varphi}/(2\,\sigma_{\varphi}^2\,\kappa)$. Substituting this result into Eq.~\eqref{cos_varphi} and Eq.~\eqref{theta1_dot}, we find  that 

\begin{equation}
    \omega_1 \equiv \langle \dot{\theta_1} \rangle \approx 
\frac{\Delta\omega}{r_{0,0}\,\Delta r_0}
 \propto \Delta\omega \label{thetadotweak1},
\end{equation}

in the small detuning limit $|\Delta\omega / \kappa|\ll \Delta r_0 \ll 1$. Here we have defined $r_{0,0} = r_0(\Delta\omega = 0)$. Furthermore, it follows from Eq.~\eqref{cos_varphi} and Eq.~\eqref{Delta r} that the slope $|\partial (\cos{\varphi_0})/\partial \Delta\omega|$ decreases as $\Delta\omega$ increases. Therefore the responsivity $\mathcal{S}$ also decreases. Qualitatively, this behavior is easily confused with an  $n^{\mathrm{th}}$ root scaling if we do not carefully consider the region where $\Delta\omega \to  0$, where Eq.~\eqref{thetadotweak1} clearly shows that the slope of the response remains finite. 

To assess the scaling of the noise floor, we linearize Eq.~\eqref{theta1_dot}, take the Fourier transform, and substitute the Green's function $G(\omega) = (i\omega - J)^{-1}$, which gives the frequency domain fluctuations in $\varphi$. This leads a general frequency noise power spectral density 

\begin{multline}
    S_{\omega_1}(\omega) \approx 
      \frac{ 2 D_{\varphi \varphi}K^{2}(\lambda_\Delta^{2}+\omega^{2}) }
            {P^2 + Q^2} \\
           - \frac{4D_{\theta_1\theta_1}K(\lambda_\Delta P+\omega Q)}
            {P^2 + Q^2}  + 2D_{\theta_1\theta_1}, 
\label{PSD_general}
\end{multline}

where $K = \kappa r_0^{-1}\sin{\varphi_0}$,  $\lambda_{\Delta} = r_0 ( \Gamma_A + \Gamma_B) - 2\sin{\varphi_0}$ (which is the restoration rate of the amplitude asymmetry factor), and
\begin{equation}
    P \equiv \lambda_{\varphi}\lambda_{\Delta} + K^2 \cos^2\varphi_0 - \omega^2, \quad
    Q \equiv \left(\lambda_{\varphi} + \lambda_{\Delta}\right)\omega. 
\end{equation}

The details of this calculation are provided in Sec.~\ref{sec:general_psd_derivation} of the Supplemental Material \cite{supplement}. Looking at the limits $\omega \to  0$, and  $|\Delta\omega/\kappa|\ll \Delta r_0 \ll 1$ gives

\begin{equation}
S_{\omega_{1}}(0)
        \approx 
          \frac{2D_{\varphi\varphi}}{(r_{0,0}\,\Delta {r_0})^{2}}, \label{S_0}
\end{equation}

which is constant with respect to $\Delta \omega$, with a form that exactly mirrors the slope in Eq.~\eqref{thetadotweak1}. To verify these findings, we simulate the Langevin equations~\eqref{Adot}-\eqref{varphi_dot} numerically with 64 independent noise realizations for each parameter set. For each spectra, we curve fit to extract the peak lasing frequency (see Sec.~\ref{sec:curve_fitting} of Supplemental Material for details \cite{supplement}). In agreement with previous studies \cite{Smith2022}, Eq.~\eqref{PSD_general} shows that the phase/frequency noise in the vicinity EP becomes colored, meaning that the uncertainty of a frequency estimate will generally depend on measurement frequency and measurement time. We address this issue by computing the average Allan deviation and fitting only the white noise tail, which yields an estimate for the white noise floor $S_{\omega_1}(0)$. This procedure was used experimentally in Ref.~\cite{Wang2020} to provide the experimental validation of Petermann-factor linewidth broadening at a linear second order EP. It is also much more data efficient than direct Monte-Carlo estimation of the frequency uncertainty. 

\begin{figure}[h]
    \centering
    \includegraphics[]{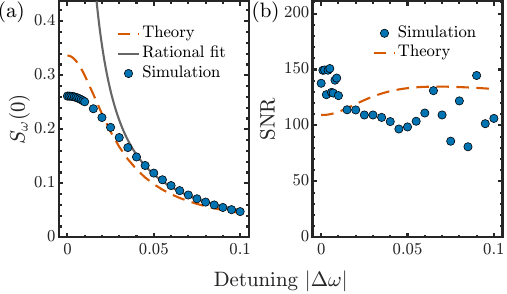}
    \caption{(a) Frequency noise floor and (b) SNR near the shifted EP. Both the theoretical Eq.~\eqref{PSD_general} and simulated results remain finite as $|\Delta\omega| \to  0$. The grey line shows a rational fit of the form $\sim |\Delta\omega|^{-4/3}$, similar to previous theory, which is accurate over a range of detuning but fails to capture the simulated noise scaling near $|\Delta\omega| = 0$. All model parameters are the same as in Fig.~\ref{fig:spectrum}.}
    \label{fig:noise}
\end{figure}

In Fig.~\ref{fig:spectrum}(c) we show that our theoretical model fits the response of the peak frequency to detuning more accurately than an $n^{\mathrm{th}}$ root scaling law when the domain extends all the way to zero detuning. We suspect that the remaining discrepancy is primarily due to the fact that amplitude noise was neglected in our analytical treatment. This result follows from the fact that the fluctuation dynamics must be considered in the analysis of the observable lasing frequency, rather than only the noise floor. In contrast, several previous studies of nonlinear EPs have calculated both the lasing frequency and the Petermann excess noise factor from the effective Hamiltonian $H_{\mathrm{eff}}$, which governs the macroscopic mean fields \cite{Bai2023a}. While some other studies have shown that the eigenmodes of the Hamiltonian governing the fluctuations must be used to predict the noise level, they still resort to $H_{\mathrm{eff}}$ to predict the frequency response \cite{Zheng2025, Smith2022}. In Ref.~\cite{Zheng2025}, this treatment predicts a square root response. However, the divergence in slope at $\Delta \omega = 0$ is not clearly observable in their numerical results. For the same parameters as in Fig.~\ref{fig:spectrum}(c), Fig.~\ref{fig:noise}(a) shows that the white noise floor saturates to a constant value as $\Delta\omega \to  0$. The later value is similar to the theoretical value in Eq.~\eqref{S_0}.  We choose to make resonator 2 entirely linear ($f_0 = 0$) for comparison with Ref.~\cite{Zheng2025}, where the authors explicitly derive a $1/\Delta \omega$ divergence in uncertainty. However, in Sec.~\ref{sec:BgD} of the Supplemental Material we show from the same Bogoliubov-de Gennes (BdG) Hamiltonian presented in their work (as opposed to our polar treatment) that a consistent treatment of second order fluctuations in this basis also leads to a finite noise floor at $\Delta\omega = 0$. Notably, the numerical results in Ref.~\cite{Zheng2025} exclude this point. Furthermore, we show that for a system with two saturable resonators ($f_0 \neq 0$) the EPs of the BdG Hamiltonian do not generally coincide with the EP of $H_{\mathrm{eff}}$; we therefore cannot rely on the divergence of noise due to mode non-orthogonality to compensate for a divergence in responsivity. 

If there is no divergence in responsivity or noise at the EP of $H_{\mathrm{eff}}$, one may ask: can these divergences be restored if we instead bias the system at an EP of $J$? From the last row of Eq.~\eqref{Jacobian_full}, the restoration rate $\lambda_{\varphi}$ of the phase fluctuation $\delta \varphi$ vanishes when $r_0 = 1$ and $\varphi_0 = \pi/2$.  In the previous analysis, only the later condition was satisfied. At this point, these dynamics clearly result in a zero eigenvalue for $J$. The symmetry of the system in Eqs.~\eqref{Adot}-\eqref{theta2_dot} under global phase translation also implies the existence of an additional zero eigenvalue. By extending $J$ into this dimension, it is straightforward to show that $r_0 = 1$, $\varphi_0 = \pi/2$ is an EP associated with the phase dimensions of the system.

To assess the response of the mean frequency in the vicinity of this EP, we begin by taking the expansion $\cos\varphi_{0}=c_{0}-s_{0}\varepsilon+O(\varepsilon^{2})$, where we define $ c_{0}\equiv\cos\varphi_{0}\big|_{\Delta\omega=0}$ and $s_{0}\equiv\sin\varphi_{0}\big|_{\Delta\omega=0}$. For the amplitude asymmetry, we obtain  $\Delta r \approx \Delta r_0 + 2\alpha_2 \varepsilon^2$, assuming that $\Delta r_0 \ll 1$. The constant $\alpha_2$ is determined from the steady-state expansion of Eqs.~\eqref{Adot}-\eqref{Bdot}. Expanding Eq.~\eqref{cos_varphi} gives the steady-state condition

 \begin{equation}
     0 = \frac{\Delta \omega}{\kappa} - \Delta r_0 c_0 + \Delta r_0 s_0 \varepsilon - c_0\alpha_2 \varepsilon^2 - 2\alpha_2 \varepsilon^3 + O(\varepsilon^4). \label{expansion_EP}
 \end{equation}

 For both the constant and linear terms of Eq.~\eqref{expansion_EP} to vanish requires $\Delta r_0 = 0$. If we also set $c_0 = 0$, the expansion states that $\cos{\varphi_0}$, and therefore also the mean frequency $\omega_1$, vary proportionally to the cubed-root of $\Delta\omega$, as in the deterministic system \cite{Bai2023a, Darcie2025}. We also see that $c_0 = 0$, $\Delta r_0 \neq 0$, which is the EP condition for $H_{\mathrm{eff}}$, is sufficient to ensure that $\omega_1$ is zeroed at $\Delta\omega = 0$, but the dominant response is still linear. This also shows that operating at the amplitude-block EP of $J$, which requires $\varphi_{0} = \pi/2, \;(\Gamma_{A}+\Gamma_{B})^{2} = 4\kappa^{2}$, is also insufficient to achieve any enhanced response. 

While operating at the phase EP seems to offer a compelling response for the deterministic system, it is problematic once fluctuations are included. Since the phase restoration vanishes directly at this EP, the phase executes pure diffusion according to $\langle\delta\varphi^{2}(t)\rangle= D_{\varphi\varphi}\,t$, and no stationary covariances exist. The initially symmetric state then drifts away from $\Delta r_0 = 0$, such that the phase restoration term is brought back at the cost of losing the divergent responsivity. To illustrate this, we combine the amplitude equations \eqref{Adot}-\eqref{Bdot} to express the Langevin equation in terms of the (now time varying) asymmetry factor $\Delta r$:

\begin{align}
\dot{\Delta r}&= 4\kappa\!\bigl[1-\sin\varphi\bigr]-\frac{\Delta \Gamma}{2}\,\Delta r\;+\;\xi_r, \label{Delta_r_dot_new}
\end{align}

where we have used $\dot{\Delta r} = \frac{1+r^{2}}{r^{2}}\;\dot r $ and defined stability margin $\Delta \Gamma \equiv \Gamma_B - \Gamma_A$, which is also related to the trace stability condition for the amplitude subsystem in Eq.~\eqref{Jacobian_full}. While this remains fully nonlinear in $\varphi$, we have dropped terms of  $\order{\Delta r^2}$ and higher in ~\eqref{Delta_r_dot_new} under the assumption that $\Delta r \ll 1$. Since we are focused on this neighborhood, we also linearize the diffusion around $r = 1$ so that the noise can be treated as additive. In the deterministic limit $\xi_r = \xi_\varphi = 0$, Eqs.~\eqref{varphi_dot} and ~\eqref{Delta_r_dot_new} have a stable fixed point at $\left(\varphi, \Delta r \right) = (\pi/2,\; 0)$, which is the location of the phase EP. To analyze the movement of this fixed point in the low noise regime, we first linearize \eqref{varphi_dot} and \eqref{Delta_r_dot_new} about $\varphi = \pi/2$. Applying It\^{o}'s lemma to $F = \Delta r$ and $F = (\delta \varphi)^2$ gives the moment equations

\begin{align}
\frac{d}{dt}\langle \Delta r \rangle &= 2\kappa \langle (\delta \varphi)^2 \rangle - \frac{\Delta\Gamma}{2}\langle \Delta r \rangle \label{moment1} \\
\frac{d}{dt}\langle (\delta \varphi)^2 \rangle &= -2\kappa \langle \Delta r (\delta \varphi)^2 \rangle + D_{\varphi\varphi}. \label{moment2}
\end{align}

For nonzero phase noise, Eq.~\eqref{moment1} predicts an upward drift in the asymmetry factor at $\langle \Delta r \rangle = 0$. When $\langle \Delta r \rangle$ is nonzero, on the other hand, $\Delta \Gamma < 0$ always leads to instability, since the phase-noise term is strictly positive. 

Furthermore, after linearizing about $\varphi=\pi/2$, the drift in $\delta\varphi$ is odd and the drift in $\Delta r$ is even with symmetric noise, so the SDEs are invariant under $\delta\varphi\to -\delta\varphi$, hence the stationary distribution is even in $\delta\varphi$. Accordingly, in the small-angle regime we adopt a joint-Gaussian ansatz for $(\Delta r,\delta\varphi)$ with $\langle\delta\varphi\rangle=0$, so the mixed moment factorizes as $\langle \Delta r\,(\delta\varphi)^2\rangle=\langle\Delta r\rangle\,\langle(\delta\varphi)^2\rangle$, which closes the two moment equations. Solving Eqs.~\eqref{moment1}-\eqref{moment2} for their stationary solutions then gives

\begin{align}
\langle(\delta\varphi)^2\rangle_{\text{st}} &= \sqrt{\frac{\Delta \Gamma D_{\varphi\varphi}}{8\,\kappa^{2}}}
\;+\; O\!\bigl(D_{\varphi\varphi}^{3/2},\,D_{rr}\bigr),\\
\langle\Delta r\rangle_{\text{st}} &= \sqrt{\frac{2\,D_{\varphi\varphi}}{\,\Delta \Gamma}}
\;+\; O\!\bigl(D_{\varphi\varphi}^{3/2},\,D_{rr}\bigr), \label{Delta_r_st}
\end{align}

 which are valid for $\Delta \Gamma > 0$. We note that with our chosen test functions only $D_{\varphi \varphi}$ emerges in our expressions; the amplitude diffusion $D_{rr}$ only enters if we track higher-order moments.  We find that maximizing the stability margin $\Delta \Gamma$ minimizes the stationary amplitude asymmetry. However, doing so also weakens the phase restoration rate, resulting in stronger phase noise.

\begin{figure}[h]
    \centering
    \includegraphics[]{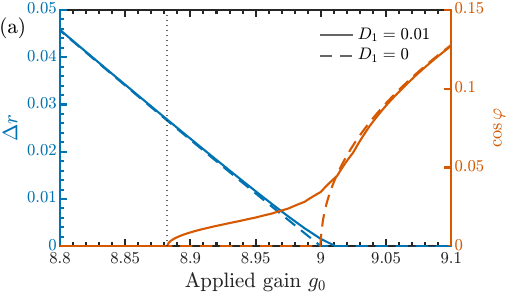}\\
    \includegraphics[]{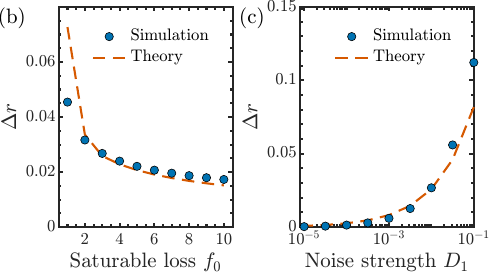}
    \caption{(a) Amplitude asymmetry factor $\Delta r$ and mean relative phase $\cos{\varphi_0}$  for different noise levels, averaged over 32 independent simulations. Since two modes lase with relative phases of $\pm\cos{\varphi_0}$ in the unbroken regime, we take $\cos{\varphi_0} \approx \Re \left(\sqrt{1 - \left(g_{\mathrm{eff}} + f_{\mathrm{eff}}\right)/4\kappa^2}\right)$. The other parameters are $f_0 = 3$, $\gamma_1 = \gamma_2 = 0.5$, $\kappa = 1$, and $D_2 = 0$. For the noiseless case (dashed lines), the exceptional point is located at $g_0 = 9$, which coincides with the point where both $\Delta r$ and $\cos\varphi_0$ are equal to zero. Once noise is introduced (solid lines), these two conditions cannot be simultaneously satisfied; the maximum $g_0$ where $\cos\varphi_0 = 0$ decreases to 8.88 (dotted line). The intersection between the dotted line and the $\Delta r$ curve is the simulated value used in (b)-(c). In (b) and (c) we repeat this process at different values of saturable loss $f_0$ and noise strength $D_1$, respectively, showing quantitative agreement between simulation results and theoretical predictions (Eq.\eqref{Delta_r_st}). We use $D_1 = 0.01$ for (b) and $f_0 = 3$ for (c), while other parameters are the same as in (a).}
    \label{fig:stability}
\end{figure}

Using Eq.~\eqref{expansion_EP}, we may now define a crossover detuning
\begin{equation}
    \Delta \omega_{\min} = \frac{\langle\Delta r\rangle_{\text{st}}^{3/2}}{\sqrt{2 \alpha_2}}, \quad  \alpha_2 =\frac{\kappa \Delta \Gamma}
      {2\bigl(\kappa^{2} - \Gamma_A\Gamma_B\bigr)},
\end{equation}

below which the response is approximately linear and the noise floor $S_{\omega_1}(0)$ is constant. To verify the result of Eq.~\eqref{Delta_r_st} we use our Monte-Carlo framework to simulate the values of $\Delta r$ and $\cos{\varphi_0}$, along with the linearized gain/loss $\Gamma_A$ and $\Gamma_B$, for a variety of parameter sets. Fig.~\ref{fig:stability}(a) shows that for nonzero noise, the zero-points in $\cos{\varphi_0}$ and $\Delta r$ occur at different values of the gain $g_0$. A similar trend can be observed in other parameter sweeps that contain the noiseless exceptional point (such as coupling or passive loss), indicating that this separation cannot be removed via parameter tuning. The minimum value of $\Delta r$ subject to $\cos{\varphi_0} = 0$ corresponds to the minimum amplitude asymmetry required to compensate for the phase noise at that point, as shown in Eq.~\eqref{moment1}-\eqref{moment2}. In Fig.~\ref{fig:stability}(b) and (c) we repeat this process at different values for the saturable loss $f_0$ and noise strength $D_1$, finding quantitative agreement between our simulated and theoretical results (Eq.~\eqref{Delta_r_st}) with no additional scaling factor. This supports the claim that this drift term prevents the system from being biased exactly at the phase EP. We note that our moment closure results are less accurate for low $f_0$ in our sweep, where the system is close to threshold, and for strong noise $D_1$, since we have neglected higher-order noise terms. 

In conclusion, we have shown that the fluctuation dynamics, rather than only the effective Hamiltonian of the mean fields, must be included to accurately describe both the responsivity and noise scaling near a nonlinear EP. Applying a self-consistent treatment of fluctuations up to second order, we find that there is no divergence in responsivity, noise, or SNR as $|\Delta\omega| \to 0$; they all saturate to constant values which are well approximated by our theoretical results. In Sec.~\ref{sec:FI} of the Supplemental Material, we also find no divergence in the Fisher information obtained for a series of phase, amplitude, or frequency measurements \cite{supplement}. Furthermore, we identify that feedback of phase noise into the amplitude subsystem pushes it away from perfect symmetry, imposing a lower limit on the proximity to the phase EP that can be attained as a steady-state operating point. While our initial simulations use a laser saturation model, these effects could also be probed experimentally using electrical circuits \cite{Bai2023a,Bai2024}, microwave cavities \cite{Dembowski2001, Chen2017}, optical ring resonators \cite{Darcie2024, Hodaei2016, Zhao2018, Peng2014}, micro-pillar cavities \cite{StJean2017}, or photonic crystal cavities \cite{Kim2016, Ji2023} with various nonlinear forms. 

For any experimental system, a truly divergent linewidth is clearly nonphysical, but would be required to reconcile the divergence in responsivity predicted by previous theory \cite{Bai2023,Bai2024,Bai2023a,Smith2022}. Due to the stringent tuning requirements for reaching an EP, residual parametric errors are easily blamed for experimentally observed limits, since they also limit the maximum responsivity \cite{Darcie2025}. However, these limits are actually intrinsic to all PT-symmetric systems (\cite{Liu2016, Hodaei2017, Mortensen2018, Xiao2019, Zeng2019, Bai2023a, Zhang2024}, for instance), which are subject to fundamental noise and saturation effects, and may also be present for EPs in systems with dissipative coupling (\cite{Wang2020, Lai2019, Hokmabadi2019}). Our findings therefore have broad implications for a wide range of possible EP-based sensors.

\begin{acknowledgments}
We would like to acknowledge financial support from the
NSERC Alliance Quantum Consortium (AQUA).
\end{acknowledgments}

\bibliography{references_updated}

\clearpage
\begin{widetext}

\makeatletter
  \setcounter{secnumdepth}{1}
  \setcounter{section}{0}    
  \renewcommand{\thesection}{S\arabic{section}}%
\makeatother

\setcounter{equation}{0}
\renewcommand{\theequation}{S\arabic{equation}}
\setcounter{figure}{0}
\renewcommand{\thefigure}{S\arabic{figure}}
\setcounter{table}{0}
\renewcommand{\thetable}{S\arabic{table}}

\begin{center}
  \textbf{Supplemental Material for}\\
  {\large “Noise-Induced Limits on Responsivity and SNR for Nonlinear Exceptional Point Sensing”}\\
  {\small Todd Darcie and J.~S.~Aitchison}
\end{center}

\section{Noise scaling and precision in the weak-noise limit}\label{sec:noise_scaling_weak}

In this section, we discuss the noise and precision scaling in the weak noise limit, which is valid when the mean-squared fluctuations in Eqs.~\eqref{Adot}-\eqref{theta2_dot} can be neglected, and the shift in the EP vanishes. With this assumption, the mean dynamics converge to the deterministic (noise-free) solution outlined in \cite{Darcie2025}, which is

\begin{equation}\label{noiseless_sol}
\begin{aligned}
\omega_{1} &\approx \sqrt{\,2\,\kappa\,\overline{b}\,}\;\bigl|\Delta\omega\bigr|^{1/3}, \\[6pt]
\varphi_{0} &\approx \frac{\pi}{2} \;-\; \frac{\sqrt{\,\kappa\,\overline{b}\,}\;\bigl|\Delta\omega\bigr|^{1/3}}{\kappa}, \\[6pt]
r_{0} &\approx 1 \;+\; \frac{\bigl|\Delta\omega\bigr|^{2/3}}{\sqrt{\,2\,\kappa\,\overline{b}\,}}, \\[6pt]
r_{0}^{-1} &\approx 1 \;-\; \frac{\bigl|\Delta\omega\bigr|^{2/3}}{\sqrt{\,2\,\kappa\,\overline{b}\,}}.
\end{aligned}
\end{equation}

where $\bar{b}$ is a closed-form expression of the parameters $f_0, \gamma_1,$ and $\gamma_2$. Close to the exceptional point, the amplitude-phase coupling for the fluctuations can be neglected, and the dynamics of the phase fluctuations can be expressed as
\begin{equation}
    \dot{x}(t) \;=\; J\,x(t) \;+\;\xi(t), \label{fluctuations_simple}
\end{equation}

where 

\begin{equation}
x(t)=\binom{\delta \theta_1(t)}{\delta \theta_2(t)}, \quad J=\left(\begin{array}{cc}
\kappa_1 & -\kappa_1 \\
\kappa_2 & -\kappa_2
\end{array}\right), \quad \xi(t)=\binom{\xi_{\theta_1}(t)}{\xi_{\theta_2}(t)} .
\end{equation}
 We have introduced constants $\kappa_1\equiv \kappa r_0^{-1} \sin{\varphi_0}$ and $\kappa_2\equiv \kappa r_0 \sin{\varphi_0}$. Taking the Fourier transform of Eq.~\eqref{fluctuations_simple}, we get 
\begin{equation}
    \widetilde{x}(\omega) 
\;=\;
\bigl[i\omega\,I - J\bigr]^{-1}\,\widetilde{\xi}(\omega).
\end{equation}

After performing this inversion, we exploit the independence of the driving noise terms to express the noise power spectral density vector $\langle|\widetilde{x}(\omega)|^2\rangle$, which has elements 

\begin{align}
 S_{\delta\theta_1}(\omega) 
= 
\frac{
2\,\bigl[
D_{\theta_1 \theta_1}\,\bigl(\omega^2 + \kappa_2^2\bigr) 
\;+\;
D_{\theta_2 \theta_2}\,\kappa_1^2
\bigr]
}{
\omega^2\,\bigl[\omega^2 + (\kappa_2 - \kappa_1)^2\bigr]
}, \\ S_{\delta\theta_2}(\omega) 
= 
\frac{
2\,\bigl[
D_{\theta_1,\theta_1}\,\kappa_2^2 
\;+\;
D_{\theta_2,\theta_2}\,\bigl(\omega^2 + \kappa_1^2\bigr)
\bigr]
}{
\omega^2\,\bigl[\omega^2 + (\kappa_2 - \kappa_1)^2\bigr]
}.
\end{align}

The relative enhancement of the noise in each resonator compared to the decoupled system ($\kappa \to  0$), is evaluated in the DC limit ($\omega \to  0$) as

\begin{align}
    K_{\theta_1} &= \frac{D_{\theta_1\theta_1}\,\kappa_2^2 + D_{\theta_2\theta_2}\,\kappa_1^2}{D_{\theta_1\theta_1}\,(\kappa_2-\kappa_1)^2},\\
    K_{\theta_2} &= \frac{D_{\theta_1\theta_1}\,\kappa_2^2 + D_{\theta_2\theta_2}\,\kappa_1^2}{D_{\theta_2\theta_2}\,(\kappa_2-\kappa_1)^2}
\end{align}

Near the EP, we can use Eq.~\eqref{noiseless_sol} to express this relative enhancement as 

\begin{align}
K_{\theta_1} &\approx \frac{2\kappa\,\bar{b}\,\bigl(D_{\theta_1\theta_1}+D_{\theta_2\theta_2}\bigr)}{D_{\theta_1\theta_1}}\Delta\omega^{-4/3}, \label{K_theta_1}\\[1mm]
K_{\theta_2} &\approx \frac{2\kappa\,\bar{b}\,\bigl(D_{\theta_1\theta_1}+D_{\theta_2\theta_2}\bigr)}{D_{\theta_2\theta_2}}\Delta\omega^{-4/3}\, \label{K_theta_2}.
\end{align}

In the best case scenario, only the $D_{\theta_1 \theta_1}$ is nonzero for $K_{\theta_1}$, and vice versa. In this case, the enhancement in noise exactly compensates for the increase in responsivity derived in \cite{Darcie2025}, yielding no precision advantage. If substantial pumping is provided to both resonators such that the spontaneous emission into both resonators is non-negligible -- and both diffusion terms contribute -- the precision of the decoupled system will exceed that of the coupled system by a constant factor. 

In Sec.~\ref{sec:petermann}, we calculate the Petermann factor of $H(t)$ directly. In this case, we find that the scaling of the noise is $\propto |\Delta\omega|^{-2/3}$, which is less deleterious than the $ |\Delta\omega|^{-4/3}$ scaling in Eqs.~\eqref{K_theta_1}-\eqref{K_theta_2}. It is also slower than the $|\Delta\omega|^{-4/3}$ scaling of the responsivity, which is given by Eq.~\eqref{noiseless_sol}. This discrepancy arises from the fact that that the Petermann factor is derived with the assumption of linearity \cite{Petermann1979}; for nonlinear systems, the Hamiltonian governing the fluctuations is different than that of the macroscopic fields. 

The results in Eqs.~\eqref{K_theta_1}-\eqref{K_theta_2} therefore show that the divergences in phase/frequency noise at the EP perfectly cancel each other, leading to a constant SNR. We note that no treatment of second order fluctuations is required to reach this conclusion. This phenomenon mirrors the behavior of linear systems, except in that that case the divergences are characterized by a slower characteristic exponent of $|\Delta\omega|^{-1}$ rather than $|\Delta\omega|^{-4/3}$ \cite{Wang2020}. 
To assess the scaling of the amplitude noise, we can apply a similar process to the amplitude fluctuations $\delta A$ and $\delta B$ which gives

\begin{align}
S_{\delta A}(\omega) &= 
\frac{
2\,D_{AA}\,\bigl[\omega^2 +\Gamma_B^2\bigr]
\;+\;
2\,D_{BB}\,\kappa^2\,\sin^2{\varphi_0}
}{
\bigl|\Delta(\omega)\bigr|^2
},
\label{eq:psdA}
\\[6pt]
S_{\delta B}(\omega) &=
\frac{
2\,D_{AA}\,\kappa^2\,\sin^2{\varphi_0}
\;+\;
2\,D_{BB}\,\bigl[\omega^2 + \Gamma_A^2\bigr]
}{
\bigl|\Delta(\omega)\bigr|^2
},
\label{eq:psdB}
\end{align}
where 
\begin{align}
\bigl|\Delta(\omega)\bigr|^2 &= 
\Bigl[
\omega^2 + \Gamma_A \Gamma_B
- \kappa^2\,\sin^2{\varphi_0}
\Bigr]^{2}
+ \,\omega^2
(\Gamma_B - \Gamma_A)^{2}.
\label{eq:DeltaSq}
\end{align}

In this case, the denominator $\Delta(0)$ is expected to be nonzero since the  gain/loss slopes in $\Gamma_A$ and $\Gamma_B$ do not need to be tuned to reach the EP at $g = f = \kappa$. Therefore, the amplitude noise does not diverge. 

\section{Petermann factor of the instantaneous Hamiltonian}\label{sec:petermann}

To compute the Petermann factor for the $H(t)$ in the weak noise limit, we will once again use the results of \cite{Darcie2025}. To leading order in $\Delta\omega$, the deviation from the noiseless EP condition $g = f = \kappa$ is 

\begin{align}
    g &= \kappa - (\bar{b} + \delta b) |\Delta\omega|^{2/3}\\
    f &= \kappa - (\bar{b} - \delta b) |\Delta\omega|^{2/3}
\end{align}
where $\bar b=\frac{\kappa}{8(\delta b ^{2})}$ is imposed by the steady-state condition. Substituting into Eq.~\eqref{H}, the normalized right-eigen-states in the vicinity of the EP are 

\begin{equation}
\bigl|\psi_{\pm}^{R}\bigr\rangle
\approx \frac{1}{\sqrt{\,2+2\bar b\,\lvert\Delta\omega\rvert^{2/3}/\kappa}}\,
\begin{pmatrix}
\,i\;\pm\;\sqrt{2\bar b/\kappa}\,\lvert\Delta\omega\rvert^{1/3}\\[6pt]
1
\end{pmatrix}
\end{equation}

to lowest order in $\Delta\omega$. Therefore, the Petermann factor is given by 

\begin{equation}
K_{\pm}\;=\;\frac{1}{\bigl|\langle\psi^{L}_{\pm}\!\mid\psi^{R}_{\pm}\rangle\bigr|^{2}}
\; \approx \frac{\kappa}{2\bar b}\;\lvert\Delta\omega\rvert^{-2/3}
\end{equation}

\section{Location of Noise Adjusted Exceptional Points}\label{sec:noise_adjusted_EPs}
In this section we outline the semi-analytical procedure for finding the shifted EP for the generalized system described in the main text. We begin by linearizing Eqs.~\eqref{Adot}-\eqref{Bdot} and \eqref{varphi_dot} by applying the substitution $S = S_0 + \delta S$ for each variable $A, B$ and $\varphi$, with $\langle \delta S \rangle = 0$. Keeping terms up to second order in the fluctuations, the steady-state mean values must satisfy 

\begin{align}
0 &= g(A_0^2)\,A_0
   + \Bigl[\,3A_0\,g'(A_0^2)+2A_0^3\,g''(A_0^2)\Bigr]\,
     \langle\delta A^2\rangle
   - \kappa\,B_0\sin\varphi_0\,
     \Bigl(1 - \tfrac12\langle\delta\varphi^2\rangle\Bigr),
\label{A_ss}\\[8pt]
0 &= -\,f(B_0^2)\,B_0
   - \Bigl[\,3B_0\,f'(B_0^2)+2B_0^3\,f''(B_0^2)\Bigr]\,
     \langle\delta B^2\rangle
   + \kappa\,A_0\sin\varphi_0\,
     \Bigl(1 - \tfrac12\langle\delta\varphi^2\rangle\Bigr),
\label{B_ss}\\[8pt]
0 &= \Delta\omega
   - \kappa\Bigl(\frac{A_0}{B_0}-\frac{B_0}{A_0}\Bigr)\cos\varphi_0\,
     \Bigl(1 - \tfrac12\langle\delta\varphi^2\rangle\Bigr)
   + \kappa\,\langle\delta A\,\delta B\rangle
     \Bigl(\frac{1}{B_0^2}-\frac{1}{A_0^2}\Bigr)\cos\varphi_0 \label{varphi_ss}
\\[-2pt]
   &\qquad
   - \kappa\Bigl(\frac{A_0\,\langle\delta B^2\rangle}{B_0^3}
                -\frac{B_0\,\langle\delta A^2\rangle}{A_0^3}\Bigr)
     \cos\varphi_0. \nonumber 
\end{align}

 Since we expect $\varphi_0 = \pi/2$ at the EP, such that the amplitude and phase dynamics are decoupled in Eqs.~\eqref{Adot}-\eqref{varphi_dot}, we have assumed for now that the cross correlations between $\delta A, \delta B$ and $\delta \varphi $ are not significant. However, we have kept correlations between $\delta A$ and $\delta B$. To first order, the dynamics of the fluctuations are

\begin{align}
\delta\dot{A} 
&= \Bigl[g(A_0^2) + 2\,A_0^2\,g'\bigl(A_0^2\bigr)\Bigr]\,\delta A
   \;-\;\kappa\,\sin\varphi_0\,\delta B
   \;+\;\kappa\,B_0\cos\varphi_0\,\delta\varphi
   \;+\;\xi_A,
\label{delta_A}\\[6pt]
\delta\dot{B} 
&= \kappa\,\sin\varphi_0\,\delta A
   \;-\;\Bigl[f(B_0^2) + 2\,B_0^2\,f'\bigl(B_0^2\bigr)\Bigr]\,\delta B
   \;-\;\kappa\,A_0\cos\varphi_0\,\delta\varphi
   \;+\;\xi_B,\label{delta_B}
\\[6pt]
\delta\dot{\varphi}
&= \kappa\,\cos\varphi_0
   \Bigl(\frac{B_0}{A_0^2} + \frac{1}{B_0}\Bigr)\,\delta A
   \;-\;\kappa\,\cos\varphi_0
   \Bigl(\frac{1}{A_0} + \frac{A_0}{B_0^2}\Bigr)\,\delta B
   \;+\;\kappa\,\sin\varphi_0
   \Bigl(\frac{B_0}{A_0} - \frac{A_0}{B_0}\Bigr)\,\delta\varphi
   \;+\;\xi_{\varphi},\label{delta_varphi}
\end{align}

We first analyze the case where $\Delta \omega = 0$. At this point, Eq.~\eqref{varphi_ss} requires that $\varphi_0  = \pi/2$, which is the same as the deterministic system \cite{Darcie2025}. We can therefore isolate the amplitude subsystem

\begin{equation}
    \dot{\mathbf u}=M_{AB}\,\mathbf u+\boldsymbol{\xi}_{AB},\qquad
\mathbf u=\!\begin{pmatrix}\delta A\\\delta B\end{pmatrix},\;
M_{AB}=
\begin{pmatrix}
\Gamma_{A}&-\kappa\\
\;\;\kappa &-\Gamma_{B}
\end{pmatrix}\!,
\end{equation}
with 
\begin{align}
\Gamma_{A}&= g(A_{0}^{2})+2A_{0}^{2}g'(A_{0}^{2}),\\
\Gamma_{B} &= f(B_{0}^{2})+2B_{0}^{2}f'(B_{0}^{2}),
\end{align}
which is stable for $
\Gamma_{A} < \Gamma_B$ and $\kappa^{2} - \Gamma_A\Gamma_B > 0$. Our previous work (Ref.~\cite{Darcie2025}) has shown that these conditions are usually satisfied at the location of an EP above threshold, provided gain relaxation rate is sufficiently fast (in this case, our laser model assumes instantaneous saturation). A unique stationary covariance matrix $\Sigma_{AB}$ can therefore be found by solving the continuous-time Lyapunov equation 

\begin{equation}
    M_{AB}\,\Sigma_{AB}+\Sigma_{AB}\,M_{AB}^{\!\top}+D=0, \label{lyap}
\end{equation}

where $D$ is the diagonal diffusion matrix with elements defined in the main text. Solving Eq.~\eqref{lyap} component-wise gives

\begin{align}
    \displaystyle 
\Sigma_{AB} &=
\begin{pmatrix}
\langle \delta A^2\rangle & \langle \delta A \delta B\rangle\\[4pt]
\langle \delta A \delta B\rangle&\langle \delta B^2\rangle
\end{pmatrix}, \label{Sigma}\\
\langle \delta A^2\rangle
&=\frac{ \bigl(\Gamma_B-\Gamma_A\bigr)\,D_{AA}\,\Gamma_B
         +\kappa^{2}\,(D_{AA}+D_{BB})}
        {2\bigl(\Gamma_B-\Gamma_A\bigr)\bigl(\kappa^{2}-\Gamma_A\Gamma_B\bigr)}, 
    \label{sigma_A}\\[8pt]
\langle \delta B^2\rangle
&=\frac{ \bigl(\Gamma_B-\Gamma_A\bigr)\,D_{BB}\,\Gamma_A
         +\kappa^{2}\,(D_{AA}+D_{BB})}
        {2\bigl(\Gamma_B-\Gamma_A\bigr)\bigl(\kappa^{2}-\Gamma_A\Gamma_B\bigr)},  \label{sigma_B}\\[8pt]
\langle \delta A \delta B\rangle
&=-\,\frac{\kappa\,(D_{AA}\Gamma_B+D_{BB}\Gamma_A)}
         {2\bigl(\Gamma_B-\Gamma_A\bigr)\bigl(\kappa^{2}-\Gamma_A\Gamma_B\bigr)}. \label{sigma_AB}
\end{align}

The Lyapunov equation for the remaining phase subsystem directly yields 

\begin{equation}
\langle\delta\varphi^2\rangle
= \frac{D_{\varphi\varphi}}
          {2\,\kappa\,\sin\varphi_0\,
           \bigl(\tfrac{A_0}{B_0}-\tfrac{B_0}{A_0}\bigr)}. \label{sigma_varphi_sm}
\end{equation}

Using a standard Taylor series expansion, the adjusted gain and loss due to these fluctuations are 

\begin{align}
g_{\mathrm{eff}} &\equiv \langle g(A^{2})\rangle
= g(A_0^{2})
+ \bigl[g'(A_0^{2}) + 2A_0^{2}g''(A_0^{2})\bigr]\;
      \langle\delta A^{2}\rangle \label{g_eff_sm}\\
      f_{\mathrm{eff}} &\equiv \langle f(A^{2})\rangle
= f(B_0^{2})
+ \bigl[f'(B_0^{2}) + 2B_0^{2}f''(B_0^{2})\bigr]\;
      \langle\delta B^{2}\rangle \label{f_eff_sm}\\
\end{align}

which form an effective Hamiltonian 

\begin{equation}
H_{\mathrm{eff}} \equiv \langle H(t) \rangle =
\begin{pmatrix}
i\,g_{\mathrm{eff}} & \kappa \\[1mm]
\kappa & -\,i\,f_{\mathrm{eff}}
\end{pmatrix}
\end{equation}

For each set of parameters, we find the self-consistent solution to Eqs.~\eqref{sigma_A}, \eqref{sigma_B}, \eqref{sigma_varphi_sm}, \eqref{A_ss}, and \eqref{B_ss}, which determines the variance terms $\langle \delta A^2 \rangle$, $\langle \delta B^2 \rangle$ along with the mean amplitudes $A_0$ and $B_0$. To reach the shifted EP, we tune the parameters of the system such that $g_{\mathrm{eff}} + f_{\mathrm{eff}} = 2\kappa$ -- the EP condition of $H_{\mathrm{eff}}$ -- at this solution. This also ensures $\varphi_0 = \pi/2$ at $\Delta\omega =0$, which is required for self-consistency. Specifically, we choose to tune the applied gain $g_0$ for the system with two saturable resonators, or the passive loss $\gamma_2$ for the system with one saturable resonator. In this later case, a closed form approximation for the location of the shifted EP can be found \cite{Zheng2025}. Since $g_{\mathrm{eff}}$ and $f_{\mathrm{eff}}$ receive different noise corrections, we no longer generally expect $A_0 = B_0$ at the shifted EP, as is the case for a deterministic EP \cite{Darcie2025}; numerical simulations at various noise strengths confirm that $B_0$ is less than $A_0$ at this point (shown by $\Delta r_0 > 0$ in Fig.~\ref{fig:stability}). 
 
\section{Curve fitting of numerical spectra}\label{sec:curve_fitting}

This section describes the approach to simulation of the system in Eqs.~\eqref{Adot}-\eqref{theta2_dot} and estimation of the peak frequencies. The additive white noise sources $\xi_{i}$ are modeled as independent Wiener processes, and the resulting SDEs are simulated using a Euler–Maruyama method \cite{Higham2001}. For each parameter set, we begin by setting the noise amplitude to zero to compute the steady-state deterministic solution, which we then use as an initial guess for 64 stochastic simulations, which each use different random noise vectors. After removing the first 20\% of the data from each run to eliminate any transient ``burn-in", we divide it into time segments, and collect Fourier spectra from each segment. To achieve a balance between resolution and smoothness, we use a run-time of $4 \times 10^4$, which is divided into 100 segments for each simulation. We first average across segment spectra to get average spectra for each simulation, which are then curve-fit independently for each simulation run. 

We can estimate for the frequency uncertainty from the standard deviation of peak frequencies extracted from each curve fit. However, this approach results in a much noisier estimate than the Allan deviation approach, which is described in Sec.~\ref{sec:allan}. 

When $\Delta\omega \neq 0$, the power spectrum is asymmetric. Furthermore, it can have two peaks, as in Fig.~\ref{fig:spectrum}(b). These features are captured by the fitting function

\begin{equation}
I(\omega)
\;=\;
\frac{\,c_{7}\,\omega^{2} \;+\; c_{6}\,\omega \;+\; c_5\,}
     {\,c_{4}\,\omega^{4} \;+\; c_{3}\,\omega^{3} \;+\; c_{2}\,\omega^{2} \;+\; c_{1}\,\omega \;+\; 1\,},  \label{fit}
\end{equation}

where $c_0,\ldots, c_7$ are fitting parameters. Example curve fits for both symmetric and asymmetric spectra are included in Fig.~\ref{fig:curve_fitting}(a) and (b), respectively. 

\begin{figure}[h]
    \centering
    \includegraphics[]{  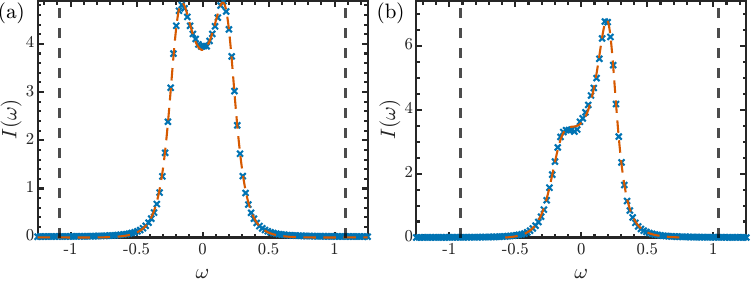}
    \caption{Example curve fits for the averaged power spectra. (a) Symmetric spectrum at the location of the deterministic EP with parameters $g_0 = 4, f_0 = 0, \gamma_1 = \gamma_2 = 1, \kappa = 1$, but with nonzero noise $D_1 = 0.01$. (b) Asymmetric spectrum with the same parameters, except $\Delta\omega = 0.005$. In both (a) and (b), the dashed orange lines show the curve fit using Eq.~\eqref{fit} and the blue crosses show the the averaged spectra obtained through our simulations. The gray vertical lines are positioned at the first and last points where $I(\omega)$ is greater than 0.1\% of its peak value, which determine the boundaries of the fitting region. }
    \label{fig:curve_fitting}
\end{figure}

\section{Power spectral density estimation}\label{sec:allan}
In this section we describe the procedure used to extract the scaling of the noise floor, which presented in Fig.~\ref{fig:noise} in the main text.  To assess the achievable sensing precision near an EP, we must estimate the  uncertainty in a measurement -- or series of measurements -- of the observable lasing frequency. This can be achieved using a Monte-Carlo approach, as described in Sec.~\ref{sec:curve_fitting}. However, this requires many independent curve fits, each with a long time span, since the noise is colored at low and intermediate frequencies. 

Following the approach used in \cite{Wang2020}, we instead calculate the average Allan deviation from the time-domain traces. The Allan deviation for a signal frequency $\nu_i = \omega_i/(2\pi)$ is defined by

\begin{equation}
\sigma_{\nu_i}(\tau) \equiv \sqrt{\frac{1}{2(M-1)} \sum_{k=1}^{M-1}\left(\bar{\nu}_{i, k+1}-\bar{\nu}_{i,k}\right)^2}
\end{equation}

where $\tau$ is the averaging time, $M$ is the number of frequency measurements, and $\bar{\nu}_{i,k}$ is the average frequency of resonator $i$ in the time interval between $k \tau$ and $(k+1) \tau$. For this we use the instantaneous frequency $\nu_{i,t} = (\theta_{i,t+dt} + \theta_{i,t})/(2\pi dt)$. The Allan deviation follows a $\tau^{-1/2}$ dependence when the underlying frequency noise spectra density is white \cite{Pikal1997}. This portion of the Allan deviation plot can be fit using $\sigma_{\nu_i} = \sqrt{S_{\omega_i}/(8\pi^2\tau)}$ where $S_{\omega_i}$ is the white frequency noise single-sided power spectral density for the noise in resonator $i$. 

In Fig.~\ref{fig:allan} we show Allan deviation plots with the same parameters as Fig.~\ref{fig:noise}(a). Going from $\Delta\omega = 0.1$ in (a) down to $\Delta\omega = 0$ in (b), the long-time Allan deviation increases, but remains finite.
With noise injected only through resonator 1, the high frequency noise (short averaging time) is white for resonator 1 (blue line), and is effectively integrated in resonator 2 (orange line), leading to random-walk frequency noise. However, for frequencies much lower than the coupling rate ($\kappa = 1$, in this case) the PSD is dominated by equivalent white noise for both resonators. The $\tau^{-1/2}$ fits are shown by the the dashed lines. In Fig.~\ref{fig:allan}(c), we also show the estimated PSD using Welche's method, which yields an estimate for $S_{\omega_i}$ that is very similar to our Allan deviation approach. 

\begin{figure}[h]
    \centering
    \includegraphics[]{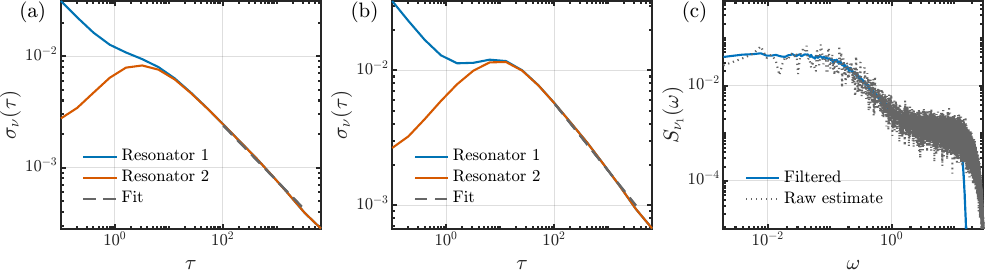}
    \caption{(a)-(b) Allan deviation measurement of frequency $(\sigma_{\nu_i}(\tau))$ versus averaging time $\tau$ for resonator $i = 1,2$, averaged over 64 independent simulations. The long-term part is fitted with $\sigma_{\nu_i} = \sqrt{S_{\omega_i}/(8\pi^2\tau)}$. We set $\Delta\omega = 0.1$ in (a) and $\Delta\omega = 0$ in (b)-(c), while the other parameters are the same as in Fig.~\ref{fig:noise}. (c) Estimated frequency noise power spectral density obtained using Welch's method with low pass filtering.}
    \label{fig:allan}
\end{figure}

\section{Frequency Noise and uncertainty near the shifted EP}\label{sec:general_psd_derivation}
 In this section, we derive the complete frequency noise power spectral density, frequency noise floor, and uncertainty for a frequency estimate near the shifted EP. To begin, we derive deterministic drift equations for the relative phase $\varphi$ and amplitude ratio $r \equiv A/B$ from Eqs.~\eqref{Adot}-\eqref{theta2_dot}. These are

 \begin{align}
     \dot\varphi&=\dot\theta_{1}-\dot\theta_{2}
           =-\Delta\omega +\kappa\Delta r\cos\varphi,\\
           \dot r&=\bigl[g(A^{2})+f(B^{2})\bigr]\,r
       -\kappa(1+r^{2})\sin\varphi .
 \end{align}

We then define fluctuations $\delta \varphi = \varphi - \varphi_0$ and $\delta r = r - r_0$. The fluctuation in the amplitude asymmetry factor is related to $\delta r$ through $\delta \Delta r = (1 + r_0^{-2})\delta r$. Applying these substitutions yields fluctuation dynamics 

\begin{equation}
    \mathbf u
=\bigl(\,\delta \varphi,\delta \Delta r,\bigr)^{\mathsf T},
\qquad
\dot{\mathbf u}=J\,\mathbf u + \Xi, \label{fluctuation_dynamics_SM}
\end{equation}

where $\Xi$ is a vector of noise variables, and the Jacobian in this reduced basis is

\begin{equation}
    J=
\begin{pmatrix}
\lambda_\varphi &
\;\;C \\[8pt]
-C A &
\lambda_\Delta
\end{pmatrix}.
\end{equation}

where we define $A = (1+r_{0}^{2})(1+r_{0}^{-2})$ and $C = \kappa \cos{\varphi_0}$. We also find that $\lambda_{\Delta} = r_0(\Gamma_A + \Gamma_B) - 2 \kappa r_0 \sin{\varphi_0}$ is the restoration rate for $\delta\Delta r$, with the coupling terms turned off. This is analogous to the phase restoration rate $\lambda_\varphi$, which is described in the main text. The diffusion matrix elements corresponding to $\Xi$ are  

\begin{equation}
D=
\begin{pmatrix}
2(D_{\theta_1}+D_{\theta_2}) & 0 \\[6pt]
0 & 2D_{\Delta\Delta}
\end{pmatrix},
\qquad
D_{\Delta\Delta}
 =(1+r_{0}^{-2})^{2}
  \frac{D_{AA}+r_{0}^{2}D_{BB}-2r_{0}D_{AB}}{B_{0}^{2}} .
\end{equation}

To analyze the noise at a given operating point $(\varphi_0, r_0)$, we need to solve Eq.~\eqref{fluctuation_dynamics_SM}. In the frequency domain, 

\begin{align}
    \tilde{\mathbf u} (\omega) &= G(\omega) \tilde{\Xi}(\omega), \label{dynamics_SM}\\
    G(\omega)&=(i\omega I-J)^{-1}.
\end{align}

Because $J$ is $2 \times 2$, $G$ can be expressed as 
\begin{equation}
G(\omega)=\frac{1}{\Delta(\omega)}
           \begin{pmatrix}
             \lambda_{\Delta}-i\omega & -C \\[6pt]
             +C A                   & \lambda_{\varphi}-i\omega
           \end{pmatrix},
\end{equation}

with

\begin{equation}
\Delta(\omega)
  =(i\omega-\lambda_{\varphi})(i\omega-\lambda_{\Delta})+C^{2}A
\end{equation}

Since $C \to  0$ near the EP, we expect that the off-diagonal terms are negligible compared to $\lambda_\Delta$. However, we keep the $C^2$ term in $\Delta(\omega)$, as it can become significant near the EP. $\lambda_\varphi$ is also expected to be small near the EP, where $\Delta r$ is minimized. In our numerical simulations, we find that these approximations hold over the parameter ranges presented in Fig.~\ref{fig:noise} in the main text. The system in Eq.~\eqref{dynamics_SM} then simplifies to 

\begin{equation}
    \tilde{\delta \varphi} (\omega) \approx G_{11}(\omega) \tilde{\xi}_{\varphi}(\omega), \quad G_{11}(\omega) = \frac{\lambda_{\Delta}-i\omega}{\Delta(\omega)},
\end{equation}

To find the PSD of the absolute frequency $\omega_1$, we define $\delta\omega_{1}(t)\equiv-\Bigl[\dot{\theta}_{1}(t)-\langle\dot{\theta}_{1}\rangle\Bigr].$ Applying the same linearization using $\delta\varphi$ and $\delta\Delta r$ as we did above to Eq.~\eqref{theta1_dot} then gives

\begin{align}
\delta\omega_1
   = - K\,\delta\varphi - \frac{C}{r_0^2 + 1}\delta \Delta r \;+\; \xi_{\theta_1}, \label{delta_theta1_SM}
\end{align}

where $K = \kappa r_0^{-1} \sin{\varphi_0}$ is expected to be much larger than $C$ near the EP. We therefore expect that $\delta\omega_1$ is primarily driven by the relative phase fluctuations $\delta \varphi$, rather than by the amplitude fluctuations $\delta\Delta r$. This is even more apparent for low frequencies $\omega \ll \lambda_\varphi$ since we expect $\lambda_\Delta \gg \lambda_\varphi$.  
If we also want to include the direct contribution of $\xi_{\theta_1}$, we must break $\xi_{\varphi}$ into its constituent parts prior to substitution into Eq.~\eqref{delta_theta1_SM}. In the frequency domain,

\begin{equation}
    \tilde{\delta\omega_1}(\omega) \approx -\left[ 1 + K G_{11}\right]\tilde{\xi}_{\theta_1} - K G_{11}\tilde{\xi}_{\theta_2}.
\end{equation}

With one-sided diffusion constants, we also use $\langle\tilde{\xi}_{\theta_{i}}\tilde{\xi}_{\theta_{i}}^{*}\rangle
   =2D_{\theta_{i}}$ to obtain 
   
\begin{align}
     S_{\omega_1}(\omega) &\approx 
      \frac{ 2 D_{\varphi \varphi}K^{2}(\lambda_\Delta^{2}+\omega^{2}) }
            {P^2 + Q^2} - \frac{4D_{\theta_1\theta_1}K(\lambda_\Delta P+\omega Q)} 
            {P^2 + Q^2}  + 2D_{\theta_1\theta_1}, \label{PSD_general_SM}\\
 P &\equiv \lambda_{\varphi}\lambda_{\Delta} + K^2 \cos^2\varphi_0 - \omega^2, \\
    Q &\equiv \left(\lambda_{\varphi} + \lambda_{\Delta}\right)\omega. 
\end{align}

From Eq.~\eqref{PSD_general_SM}, we note that frequency noise, which is fed with white phase noises $\xi_{\theta_1}$ and $\xi_{\theta_2}$, is colored through the interaction with our system. This coloring has been previously theorized for nonlinear EPs \cite{Smith2022}, and experimentally demonstrated for other systems with non-orthogonal lasing modes \cite{Lee1998, Lee2000}. We also note that the first two terms go to zero if the two resonators are decoupled by setting $\kappa$ to zero, resulting in a trivial single-resonator power spectral density of $2 D_{\theta_1 \theta_2}$.

In assessing measurement precision, the long averaging time ( low $ \omega$) tail of the noise PSD is of paramount interest. In this region, $S_{\omega_1}$ is dominated by white frequency noise. In our case, we are also concerned with the scaling of this noise floor in the vicinity of the shifted EP at $\Delta\omega = 0$. We therefore choose to take the limits $|\Delta\omega/\kappa|\ll \Delta r_0 \ll 1$ and $\omega \to  0$. Equation Eq.~\eqref{PSD_general} then simplifies to

\begin{equation}
S_{\omega_{1}}(0)
        =2D_{\theta_1\theta_1}
          \left(
              1+\frac{2}{r_{0,0}\,\Delta r_{0}}
          \right)
          +\frac{2D_{\varphi\varphi}}{(r_{0,0}\,\Delta r_{0})^{2}},
\end{equation}

For $\Delta r_0 \ll 1$ the last term is dominant. However, we find that the contribution of the other terms con become non-negligible for $D \gtrsim 0.01$. While we have focused our analysis on $\omega_1$, we note that $\langle \omega_1 \rangle = \langle \omega_2 \rangle$, since $\varphi_0$ is constant in steady-state. While the full PSDs associated with $\delta\omega_1$ and $\delta\omega_2$ have slightly different forms, they share the same low-frequency white noise floor. This is confirmed through the Allan deviation analysis in Sec.~\ref{sec:allan}. 

For comparison with previous results in \cite{Zheng2025}, we note that the RMS frequency uncertainty over a sufficiently long averaging time $\tau$ is

\begin{equation}
    \sigma_{\omega_1}(\tau)
=\sqrt{\frac{S_{\omega_1}(0)}{2\,\tau}}. \label{sigma_omega1}
\end{equation}

\section{Full linearization using Bogoliubov-de Gennes Hamiltonian}\label{sec:BgD}

Recent work by Zheng et al. predicted a divergence of the noise at the exceptional point. They reached their conclusions by analyzing the properties of the Bogoliubov-de Gennes Hamiltonian, which simultaneously includes dynamics of the full linearized state.  In this section, we repeat this analysis for the more general case of two coupled nonlinear resonators subject to saturable gain and absorption. We find that the exceptional point of the effective Hamiltonian for the macroscopic observables only coincides an exceptional point in this linearized basis under specific circumstances. Specifically, they coincide when one resonator is entirely passive. Even in this case, however, we find the noise still saturates to a finite value if terms up to second order in the fluctuations are properly included throughout. Following the analysis of Zheng et al., we begin with the Langevin equations

\begin{align}
\dot{\psi}_1 &= \left[-i(\omega_0 + \Delta\omega) + g(|\psi_1|)\right]\psi_1 - i\kappa\,\psi_2 + \xi_{\psi_1}(t), \label{langevin1}\\
\dot{\psi}_2 &= \left[-i\omega_0 - f(|\psi_2|)\right]\psi_2 - i\kappa\,\psi_1 + \xi_{\psi_2}(t), \label{langeven2}
\end{align}

and then take the expansion

\begin{equation}
\begin{aligned}
\psi_1 &= A_0 e^{-i \omega_0 t} + \delta \psi_1\\
\psi_2 &= B_0 e^{-i \omega_0 t + i \varphi_0} + \delta \psi_2.
\end{aligned}
\end{equation}

Substituting into Eqs.~\eqref{langevin1}-\eqref{langeven2} yields

\begin{align}
\dot{\delta\psi}_1 &= \Bigl[-i(\omega_0+\Delta\omega) + g(A_0^2)\Bigr]\delta\psi_1 
+ g'(A_0^2)A_0^2\Bigl[\delta\psi_1 + e^{-2i\omega_0t}\,\delta\psi_1^*\Bigr] \nonumber \\
&\quad\quad - i\kappa\,\delta\psi_2 + \xi_{\psi_1}(t), \\
\dot{\delta\psi}_2 &= \Bigl[-i\omega_0 -f(B_0^2)\Bigr]\delta\psi_2 
- f'(B_0^2)B_0^2\Bigl[\delta\psi_2 + e^{-2i\omega_0t+2i\varphi_0}\,\delta\psi_2^*\Bigr] \nonumber \\
&\quad\quad - i\kappa\,\delta\psi_1 + \xi_{\psi_2}(t),
\end{align}

where we have neglected higher-order fluctuations. We then transform into a rotating frame at frequency $\omega_0$, and express the coefficients of $\delta \psi_1$ and $\delta \psi_2$ as $g_{\mathrm{eff}}$ and $f_{\mathrm{eff}}$ at the shifted EP, which gives

\begin{align}
\dot{\delta\psi}_1 &= \Bigl[-i\Delta\omega + g_{\mathrm{eff}} + g'(A_0^2)A_0^2\Bigr] \delta\psi_1 
+ g'(A_0^2)A_0^2\,\delta\psi_1^*
- i\kappa\,\delta\psi_2 + \xi_{\psi_1}(t), \\
\dot{\delta\psi}_2 &= -\Bigl[f_{\mathrm{eff}}  + f'(B_0^2)B_0^2\Bigr] \delta\psi_2 
- f'(B_0^2)B_0^2\,\delta\psi_2^*
- i\kappa\,\delta\psi_1 + \xi_{\psi_2}(t).
\end{align}

We can then express the dynamics of the perturbations in the basis $|\Psi\rangle = \left[
\delta \psi_1,\, \delta \psi_2,\, \delta \psi_1^*,\, \delta \psi_2^*
\right]^{T}$ as

\begin{equation}
    i\frac{d}{dt}|\Psi\rangle = \boldsymbol{H}_B\,|\Psi\rangle + \left| \xi \right),
\end{equation}

where 

\begin{align}
\boldsymbol{H}_B &= 
\begin{pmatrix}
\boldsymbol{H}_0+\boldsymbol{V}_0 & \boldsymbol{V}_0 \\[1mm]
-\boldsymbol{V}_0^*            & -\boldsymbol{H}_0^*-\boldsymbol{V}_0^*
\end{pmatrix}, \label{H_B_bloc}\\
\boldsymbol{H}_0 &= 
\begin{pmatrix}
\Delta\omega + i\,g_{\mathrm{eff}} & \kappa \\[1mm]
\kappa & -\,i\,f_{\mathrm{eff}}
\end{pmatrix}, \\
\boldsymbol{V}_0 &= 
\begin{pmatrix}
i\,g'(A_0^2)A_0^2 & 0 \\[1mm]
0 & -\,i\,f'(B_0^2)B_0^2 
\end{pmatrix}, \\
\left| \xi \right) &= 
\left[ \xi_{\psi_1}, \xi_{\psi_2}, \xi_{\psi_1}^*, \xi_{\psi_2}^* \right]^T.
\end{align}

We expect that the location of the EP of the effective Hamiltonian $\boldsymbol{H}_0$ is $\Delta \omega = 0$ and $g_{\mathrm{eff}} + f_{\mathrm{eff}}  = 2 \kappa$. At this point, the eigenvalues of the Bogoliubov-de Gennes Hamiltonian $\boldsymbol{H}_B$ are 

\begin{align}
\lambda_{1,2} &= \frac{i}{2}\Bigl(g_{\mathrm{eff}}-f_{\mathrm{eff}}\Bigr) + i\,A_0^2\,g' \;\pm\; \frac{1}{2}\sqrt{\,4\kappa^2 - \Bigl[2A_0^2\,g' + f_{\mathrm{eff}}+g_{\mathrm{eff}}\Bigr]^2},\\[1mm]
\lambda_{3,4} &= \frac{i}{2}\Bigl(g_{\mathrm{eff}}-f_{\mathrm{eff}}\Bigr) - i\,B_0^2\,f' \;\pm\; \frac{1}{2}\sqrt{\,4\kappa^2 - \Bigl[2B_0^2\,f' + f_{\mathrm{eff}}+g_{\mathrm{eff}}\Bigr]^2}. \label{H_B_val_3}
\end{align}

The corresponding right eigenstates are

\begin{align}
|\psi^R_{1,2}\rangle&= N_{1,2}
\begin{pmatrix}
\displaystyle -\frac{i\Bigl(\Gamma_{\mathrm{eff}}+2A_0^2g'\Bigr) \pm \sqrt{4\kappa^2-\Bigl(\Gamma_{\mathrm{eff}}+2A_0^2g'\Bigr)^2}}{2\kappa}\\[2ex]
-1\\[2ex]
\displaystyle -\frac{i\Bigl(\Gamma_{\mathrm{eff}}+2A_0^2g'\Bigr) \pm \sqrt{4\kappa^2-\Bigl(\Gamma_{\mathrm{eff}}+2A_0^2g'\Bigr)^2}}{2\kappa}\\[2ex]
1
\end{pmatrix}\label{H_B_vec_1},\\
|\psi^R_{3,4}\rangle &= N_{3,4}
\begin{pmatrix}
\displaystyle \frac{i\Bigl(\Gamma_{\mathrm{eff}}+2B_0^2f'\Bigr) \pm \sqrt{4\kappa^2-\Bigl(\Gamma_{\mathrm{eff}}+2B_0^2f'\Bigr)^2}}{2\kappa}\\[2ex]
1\\[2ex]
\displaystyle -\frac{i\Bigl(\Gamma_{\mathrm{eff}}+2B_0^2f'\Bigr) \pm \sqrt{4\kappa^2-\Bigl(\Gamma_{\mathrm{eff}}+2B_0^2f'\Bigr)^2}}{2\kappa}\\[2ex]
1
\end{pmatrix}, \label{H_B_vec_3}
\end{align}

where $\Gamma_{\mathrm{eff}} \equiv g_{\mathrm{eff}} + f_{\mathrm{eff}}$ is the effective gain-loss contrast and $N_{i}$ are normalization constants. 
We note that $\boldsymbol{H}_B$ has a second order EP at $g_{\mathrm{eff}} + f_{\mathrm{eff}}  = 2 \kappa$, as is the case for $\boldsymbol{H}_0$, only in the special case where the second resonator is entirely passive ($f' = 0)$. Otherwise, there are two possible EPs, both with $g_{\mathrm{eff}} + f_{\mathrm{eff}}  \neq  2 \kappa$. 

This result also contradicts previous results for linear EPs ($g' = f' = 0$), where the effective gain/loss experienced by small fluctuations are equivalent to those experienced by the full fields. In this case, $\boldsymbol{H}_B$ and $\boldsymbol{H}_0$ share the same manifold, so that they reach the same EP. 

If the EPs of $\boldsymbol{H}_B$ and $\boldsymbol{H}_0$ are different, approaching the EP of $\boldsymbol{H}_0$ will still result in nondegenerate eigenstates for $\boldsymbol{H}_B$, as described by Eqs.~\eqref{H_B_vec_1}-\eqref{H_B_vec_3}. We therefore expect that the noise will saturate to some finite value due to the non-divergent Petermann factor for these modes. 

For this reason, we focus our subsequent analysis on the EPs of $\boldsymbol{H}_B$ to assess whether there is any operating point $f_{\mathrm{eff}}$, $g_{\mathrm{eff}}$ that does result in divergent noise scaling. Interestingly, we see that both EPs of  $\boldsymbol{H}_B$ could be reached simultaneously if $2A_0^2\,g' + f_{\mathrm{eff}}+g_{\mathrm{eff}} = 2B_0^2\,f' + f_{\mathrm{eff}}+g_{\mathrm{eff}} = 2\kappa$. However, the eigenstates $|\psi^R_{1,2}\rangle$ and $|\psi^R_{3,4}\rangle$ remain distinct, meaning that this would not constitute a fourth-order EP. 

Since the applied gain $g_0$ must exceed the absorption $f_0$ above threshold, and consequentially $A_0 > B_0$, we expect that $A_0^2 g'$ is always greater in magnitude than $B_0^2 f'$. These EP conditions should therefore be distinct. Furthermore, this means that the magnitude of the parametric adjustment needed to reach the second EP (between modes 3 and 4) is expected to be lower.

To weigh the relative importance of each set of modes, we note that the derivative terms contribute differently to their respective EPs. For $f_0 > 0$ we get that $B_0^2 f', A_0^2 g' < 0$, which leads to a positive imaginary contribution for $\lambda_{3,4}$ and a negative contribution for $\lambda_{1,2}$. If we have $f_0 < 0$ and $|f_0| < g_0$ the imaginary contribution to $\lambda_{3,4}$ is still less negative. If neither of the above conditions for $f_0$ hold, the resonators could simply be swapped such that $g_0 \leftrightarrow -f_0$ and $\gamma_1 \leftrightarrow \gamma_2$.

When biased at this second EP, Eqs.~\eqref{H_B_val_3}-\eqref{H_B_vec_3} simplify to 
\begin{align}
\lambda_{EP} = i \gamma_0, \quad
|\psi^R_{EP}\rangle = \frac{1}{2}
\begin{pmatrix}
 i\\
1\\
-i\\
1
\end{pmatrix},
\end{align}

where $\gamma_0 \equiv g_{\mathrm{eff}} - \kappa$. For $\Delta \omega \neq 0$, we assume that the effective gain and loss saturate according to 

\begin{align}
    g_{\mathrm{eff}} &= g_{\mathrm{eff}} (\Delta\omega = 0) - (\bar{\chi} + \delta\chi) \Delta\omega^2 \\
    f_{\mathrm{eff}} &= f_{\mathrm{eff}} (\Delta\omega = 0) - (\bar{\chi} - \delta\chi) \Delta\omega^2 \\
\end{align}

when the noise us strong compared to the detuning, since the mean amplitude ratio $r_0$ also varies quadratically in $\Delta\omega$. From here, we calculate the splitting of the degenerate eigenvalues and eigenstates using degenerate perturbation theory \cite{Seyranian2003}. The results are 

\begin{align}
\lambda_{\pm} &= \lambda_{\mathrm{EP}} \pm \eta\,\lambda_2, \label{vals_eps}\\[8pt]
\bigl|\psi^R_{\pm}\bigr\rangle &= 
\begin{pmatrix}
\frac{i}{2}\\[4pt]
\frac12\\[4pt]
-\frac{i}{2}\\[4pt]
\frac12
\end{pmatrix}
+\eta
\begin{pmatrix}
\displaystyle \frac{\kappa(\kappa-2\beta)}{4D}\;\pm\;\frac{\lambda_2}{2\kappa}\\[8pt]
\displaystyle -\frac{i\,\kappa^2}{4D}\\[8pt]
\displaystyle \frac{\kappa(\kappa-2\beta)}{4D}\;\mp\;\frac{\lambda_2}{2\kappa}\\[8pt]
\displaystyle \frac{i\,\kappa^2}{4D}
\end{pmatrix},\label{vecs_R_plus}\\[12pt]
\bigl\langle\psi^L_{\pm}\bigr| &=
\begin{bmatrix}
\kappa & -i\kappa & -\kappa & -i\kappa
\end{bmatrix}
+\eta
\begin{bmatrix}
\displaystyle \frac{i\,\kappa^2\,(2\beta-\kappa)}{2D}, &
\displaystyle -\frac{\kappa^3}{2D}\;\pm\;\lambda_2, &
\displaystyle \frac{i\,\kappa^2\,(2\beta-\kappa)}{2D}, &
\displaystyle \frac{\kappa^3}{2D}\;\pm\;\lambda_2
\end{bmatrix}, \label{vecs_L_plus}
\end{align}
where 
\begin{align}
    \lambda_2
&= \kappa^{3/2}
\sqrt{\frac{2\beta-\kappa}{2\,D}\;-\;\bar{\chi}}\,.\\  
    D &= 2\alpha\beta - \alpha\kappa + \beta\kappa, \\[8pt]
    \langle\psi^L_{\pm}|\psi^R_{\pm}\rangle
&\approx  \pm 2\,\eta\,\lambda_2. \label{K_eps}
\end{align}

We have also introduced the shorthand $\eta = \Delta\omega/\kappa$, $ \alpha = A_0^2 g'(A_0^2)$,  $\beta = B_0^2 f'(B_0^2) $. Since $\alpha$ and $\beta$ are negative and $\kappa$ and $\bar{\chi}$ are positive, we expect that $\lambda_2$ is imaginary. We therefore express it as $\lambda_2 = i \Lambda$, with $\Lambda \in \mathbb{R}$. Considering only the two nearly-degenerate modes, we approximate the Green's function $\mathbf{G}(\omega)$ as 

\begin{equation}
\mathbf{G}(\omega) \approx 
\frac{1}{\omega - \lambda_+} 
\frac{|\psi_+^R\rangle \langle \psi_+^L|}{\langle \psi_+^L | \psi_+^R \rangle}
+ 
\frac{1}{\omega - \lambda_-} 
\frac{|\psi_-^R\rangle \langle \psi_-^L|}{\langle \psi_-^L | \psi_-^R \rangle}.
\end{equation}

To get the phase fluctuations of the field $\psi_1$, we construct a projector $P = \frac{1}{2i A_0} \begin{pmatrix} 1 & 0 & -1 & 0 \end{pmatrix}\,$ such that 

\begin{align}
\delta \theta_1 &= -\,i\,\frac{\delta \psi_1 - \delta \psi_1^*}{2A_0}
\;=\;P\,\ket{\Psi}\,.
\end{align}

 The Green's function for the phase fluctuations relates to the phase noise PSD through 

 \begin{align}
S_{\theta_1}(\omega)
&=\sum_{k=1}^4 D_k
\biggl|\sum_{\sigma=\pm}
\frac{\;\bigl\langle  P\,\bigm|\psi_\sigma^R\bigr\rangle\,
                \bigl\langle\psi_\sigma^L\bigm|e_k\bigr\rangle}
     {\omega - \lambda_\sigma}
\biggr|^2,
 \end{align}

where $e_k$ are the canonical basis vectors of $|\Psi \rangle$. Evaluating the summation gives

\begin{align}
S_{\theta_1}(\omega)
&= \frac{D_{\psi_1\psi_1}\bigl((\gamma_0+\kappa)^2 + \omega^2\bigr)
         + D_{\psi_2\psi_2}\,\kappa^2}
        {2\,A_0^2\Bigl[
            \omega^4
            + 2\bigl(\gamma_0^2 + \eta^2\Lambda^2\bigr)\,\omega^2
            + \bigl(\gamma_0^2 - \eta^2\Lambda^2\bigr)^2
         \Bigr]},
\end{align}

which is related to the frequency noise PSD through $S_{\omega_1}(\omega) = \omega^2 S_{\theta_1} (\omega)$. 
We note that the magnitude of $\gamma_0$  scales with the magnitude of the fluctuations, approaching zero as $g_{\mathrm{eff}} \to  g$. Similarly, the expansion in Eqs.~\eqref{vals_eps}-\eqref{K_eps} is valid for $\eta << 1$, so we assume $\eta \lambda_2 << \kappa$. This noise scaling that results from these assumptions is outlined in the following table: 

\begin{table}[h]
\centering
\setlength{\tabcolsep}{12pt}
\begin{tabular}{l l c c l}
\toprule
Regime & Condition on \(\omega\) & $S_{\omega_1}\propto\omega^s$ & \(\sigma_y(\tau)\propto\tau^\mu\) & Noise type \\
\midrule
I   & \(\omega \ll \omega_1\)              & \(+2\)   & \(-1\)    & White PM       \\
II  & \(\omega_1 \ll \omega \ll \omega_2\) & \(0\)    & \(-\tfrac12\) & White FM(I)   \\
III & \(\omega_2 \ll \omega \ll \kappa\)   & \(-2\)   & \(+\tfrac12\) & RW FM          \\
IV  & \(\omega \gg \kappa\)               & \(0\)    & \(-\tfrac12\) & White FM (II)  \\
\bottomrule
\end{tabular}
\end{table}

where 

\begin{equation}
    \omega_1 = \frac{\bigl|\gamma_0^2 - \eta^2\Lambda^2\bigr|}{\omega_2},
\qquad
\omega_2 = \sqrt{2\,(\gamma_0^2 + \eta^2\Lambda^2)}.
\end{equation}

For $\eta \Lambda \ll \gamma_0$, we get that $\omega_2 = 2 \omega_1$. As $\eta \Lambda$ grows toward $\gamma_0$, we note that $\omega_1$ goes to zero, so region II widens as region I shrinks. Due to the $\omega^2$ scaling of $S_{\omega_1}$ in region I, we also expect the most significant contribution to the low frequency error to occur close to region II. 

For these reasons, we expect the error for long measurement time to be dominated by the white frequency modulation in region II, which is 
\begin{align}
    S_{\omega_1}(\omega)
&\approx \frac{\bigl(D_{\psi_1\psi_1}+D_{\psi_2\psi_2}\bigr)\,\kappa^2}
     {4\,A_0^2\,\bigl(\gamma_0^2+\eta^2\Lambda^2\bigr)}
\quad(\omega_1\ll\omega\ll\omega_2). \label{S_omega1}
\end{align}

For this white-FM noise plateau, the RMS frequency uncertainty over an averaging time $\tau$ is given by Eq.~\ref{sigma_omega1}. If we set $\gamma_0 = 0$, this yields the same result as Zheng et al. for the case where $f' = 0$. However, $\gamma_0$ is only zero if $g_{\mathrm{eff}} = \kappa$, which is valid to first order in the fluctuations, but is violated once second order fluctuations are accounted for, as shown in Eq.~\eqref{g_eff_sm}. To lowest order in $\eta$, the Petermann factor for the eigenmodes of $\mathbf{H_B}$ can be equivalently computed using the $+$ or $-$ terms in Eqs.~\eqref{vecs_L_plus}-\eqref{vecs_R_plus} as

\begin{equation}
K \equiv \frac{1}{\bigl|\langle \psi^L \mid \psi^R \rangle\bigr|^2} \approx \frac{\kappa^2}{\Lambda^2 \Delta\omega^2},
\end{equation}

which predicts an uncertainty scaling of $K^{1/2} \propto \Delta\omega^{-1}$. However, from Eqs.~\eqref{S_omega1}-\eqref{sigma_omega1}, we find that for nonzero $\gamma_0$, the divergence of $\sigma_{\omega_1}$ as we approach the shifted EP is always slower than $\Delta\omega ^{-1}$. Notably, Zheng et al. observe a scaling of $\Delta\omega ^{-0.94}$, while excluding the region from $\Delta\omega = 0$ to $\Delta\omega \sim 0.01$. For $\Delta \omega$ approaching zero, the noise remains finite, with a magnitude determined by $\gamma_0$.

In Fig.~\ref{fig:BdG} we compare out updated theory, expressed in Eq.~\eqref{S_omega1}, with the results of Zheng et al \cite{Zheng2025}. As with our previous polar theory (Fig.~\ref{fig:noise}, Section \ref{sec:general_psd_derivation}), our updated BdG theory demonstrates quantitative agreement with simulation results, without any fitting parameters. Residual errors in the theory could be due to the small detuning approximation being imprecise at $\Delta\omega = 0.1$, or due to the fact that we neglected higher order nonlinear corrections, amplitude-phase cross correlations, and higher order statistical moments. We see that the previous theory fails to capture the saturation of the noise level as $\Delta\omega \to 0$. This discrepancy is due to the fact that the $K$ calculated from the eigenstates of the linear system $\mathbf{H_B}$ describes only the non-orthogonality of fluctuation modes, not the non-orthogonality of the cavity eigenmodes. Since this full system is inherently nonlinear, computing a Petermann correction using these modes would also be insufficient, as we showed in the weak noise limit. 

\begin{figure}[h]
    \centering
    \includegraphics[]{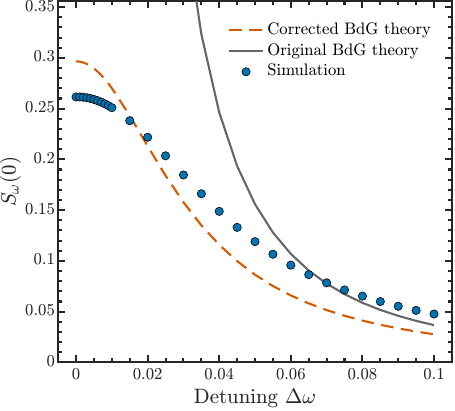}
\caption{Comparison of frequency noise $S_\omega(\omega)$ near the shifted EP computed using numerical simulations (blue circles) with frequency noise calculated from the BdG Hamiltonian $\mathbf{H_B}$, through both our analysis and using the original theory. The orange-dashes show the results of our calculation (Eq.~\eqref{S_omega1}), and the black line shows the result of \cite{Zheng2025}, where the authors assume $\gamma_0 = 0$. All model parameters are the same as in Fig.~\ref{fig:noise}. }
    \label{fig:BdG}
\end{figure}

\section{Fisher information and Cramér–Rao lower bound for frequency estimation}\label{sec:FI}

Our SNR calculations in the main text found no divergence in SNR at the location of the shifted EP, assuming a measurement scheme where the output frequency $\omega_1$ is measured over some time interval. However, this may not be the best possible measurement scheme with which to estimate $\Delta\omega$. In this section, we outline our procedure for converting simulated trajectories into numerical estimates of the Fisher information for $\Delta\omega$, which give the Cramér–Rao lower bound on the precision of the optimal estimator, given access to the complete state $X(t)=\!\begin{bmatrix}A(t),B(t), \theta_1(t), \theta_2(t)\end{bmatrix}^{\top}$. Starting from the Euler-Maruyama update for each step in our simulation 
\begin{equation}
    X_{k+1}=X_k+F(X_k, \Delta\omega)\,dt+G(X_k, \Delta\omega)\sqrt{dt}\;\xi_k ,
\qquad
\xi_k\sim\mathcal N(0,I),
\end{equation}

we treat the interval $\Delta t \equiv N_s dt$ between successive recorded samples as a short-time propagator who's statistics can be obtained by iterating this step $N_s$ times. For this section, we use $N_s = 10$. At sub-step $j$ we evaluate the Jacobian $J_j=\partial_XF(X_j,\Delta\omega)$ and write the local state-transition matrix $A_j=\partial_XF(X_j,\Delta\omega)$. The deterministic drift is accumulated over each sub-step by the Euler sum 

\begin{equation}
    \mu_{j+1}= \mu_j+F(\mu_j,\Delta\omega)\,dt,
\qquad
\mu_0=X_k ,
\end{equation}

while the covariance is accumulated by 

\begin{equation}
\Sigma_{j+1} = \Phi_j \Sigma_j \Phi_j^{\top} + G_j G_j^{\top} dt, \qquad \Sigma_0 = 0.
\end{equation}

After $N_s$ such sub-steps we obtain the effective drift $\mu_{k+1|k}\equiv\mu_{N_s}$ and effective covariance $\Sigma_{k+1|k}\equiv\Sigma_{N_s}$ that connect $X_k$ to $X_{k+1}$ over the recorded interval. Assuming that higher-order moments remain negligible over this interval, the transition density is taken to be multivariate normal:

\begin{equation}
p\left(X_{k+1} \mid X_k; \Delta\omega\right) = \mathcal{N}\left(X_{k+1}; \mu_{k+1|k}, \Sigma_{k+1|k}\right).
\end{equation}

When only a portion of the complete state is measured, we construct a projection matrix $P$ (rows of the identity that correspond to the observed variables) and replace $(\mu,\Sigma)\mapsto(P\mu,\,P\Sigma P^{\top})$. The total log-likelihood for each trajectory of $M$ recorded steps and $N_{tr}$ independent trials is 
\begin{equation}
    \ell(\Delta\omega)=\sum_{i=1}^{N_{\mathrm{tr}}}\sum_{k=0}^{M-1}
\log\mathcal N\!\bigl(PX_{k+1}^{(i)};P\mu_{k+1|k}^{(i)},\,P\Sigma_{k+1|k}^{(i)}P^{\top}\bigr),
\end{equation}

from which we find that the maximum-likelihood estimate $\widehat{\Delta\omega}$ is equal to the chosen value $\Delta\omega$, which is a simulation parameter. The Fisher information is then 

\begin{equation}
I_{\rm obs}(\widehat{\Delta\omega})
\;=\;
\left.-\,\frac{\partial^2}{\partial\Delta\omega^2}\,\mathcal L(\Delta\omega)\right|_{\Delta\omega=\widehat{\Delta\omega}}, \label{I_obs}
\end{equation}

which we approximate using a centered finite difference. The Cramér–Rao lower bound for unbiased estimation of $\Delta\omega$ is then

\begin{equation}
    \operatorname{Var}\bigl[\widehat{\Delta\omega}\bigr]\;\ge\;1/I_{\text{obs}}(\widehat{\Delta\omega}).
\end{equation}

This procedure is exact for linear SDEs and accurate for weakly nonlinear dynamics. The Fisher information curves obtained from the numerical procedure ---using Eq.~\eqref{I_obs} --- are compared with the single--resonator benchmark in Fig.~\ref{fig:FI}.  Panel~(a) sweeps the coupling rate $\kappa$.  As $\kappa\to 0$ the two resonators decouple, and the numerical result converges exactly to the single-resonator formula $I_{\theta_1}=T/D_{\theta_1}$ (gray dashed line).  Increasing $\kappa$ --- which moves the system into the unbroken symmetry regime --- transfers some $\Delta\omega$--sensitivity into the amplitudes, but no special enhancement is observed at the EP (dotted gray line).  Panels~(b) and~(c) corroborate this by perturbing, respectively, the passive loss $\gamma_2$ and the detuning $\Delta\omega$ away from the shifted EP.  In all cases the relative phase carries far more information than either amplitude or absolute phase, and the total Fisher information remains finite, smooth, and non‐extremal as the system crosses the EP. These properties also apply to both the theoretical and simulated SNR values in Fig.~\ref{fig:noise}(b).

\begin{figure}[h]
    \centering
    \includegraphics[]{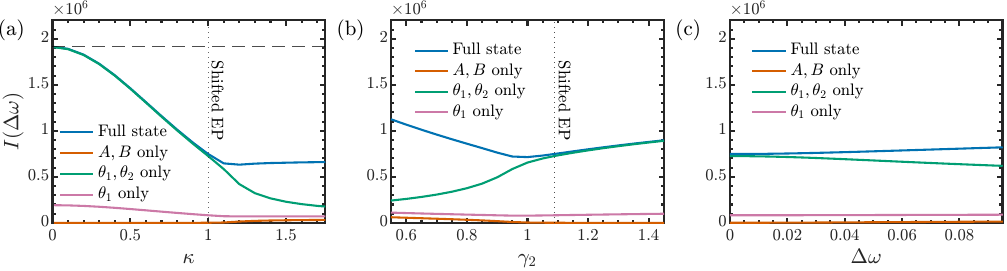}
    \caption{Numerical computation of classical Fisher information $I$ for estimation of detuning $\Delta\omega$ for a measurement period $T = 4\times 10^3$. For (a), we vary the coupling strength $\kappa$, while other parameters are the same as Fig.~\ref{fig:noise}. When $\kappa \to 0$, the numerical result (blue solid) coincides with analytic single-resonator value $I(\Delta\omega)=T/D_{\theta_1}$ (dashed gray). (b)~Passive-loss sweep through the shifted EP (gray dotted). (c)~Detuning sweep through the shifted EP (\(\Delta\omega=0\)). In all panels, we calculate $I(\Delta\omega)$ while observing both the full state (solid blue) as well as select subsets of the state variables, while masking the other non-observed variables.}
    \label{fig:FI}
\end{figure}

\end{widetext}
\end{document}